\documentclass[10pt]{iopart}
\usepackage{iopams}
\usepackage{amstext}
\usepackage{cite}
\usepackage{graphicx}
\usepackage{multicol}
\usepackage{color}

\usepackage[normalem]{ulem}

\begin{document}

%\title[Analyses of the Coulombic physics of and global spin twists in the HXY model]{Analyses of the approximate Coulombic physics of and global spin twists in the two-dimensional harmonic XY model}

\title{An electric-field representation of the harmonic XY model}

\author{Michael F. Faulkner$^{1,2,3}$, Steven T. Bramwell$^1$ and Peter C. W. Holdsworth$^2$}

\address{$^1$ London Centre for Nanotechnology and Department of Physics and Astronomy, University College London, 17-19 Gordon Street, London WC1H 0AH, United Kingdom}
\address{$^2$ Laboratoire de Physique, CNRS UMR 5672, \'{E}cole Normale Sup\'{e}rieure de Lyon, Universit\'{e} de Lyon, 46 all\'{e}e d'Italie, F-69342 Lyon Cedex 07, France}
\address{$^3$ School of Mathematics, University of Bristol, University Walk, Bristol BS8 1TW, United Kingdom}

\eads{\mailto{michael.faulkner@bristol.ac.uk}}

%\vspace{10pt}
%\begin{indented}
%\item[]February 2014
%\end{indented}

\begin{abstract}
The two-dimensional harmonic XY (HXY) model is a spin model in which the classical spins interact via a piecewise parabolic potential. We argue that the HXY model should be regarded as the canonical classical lattice spin model of phase fluctuations in two-dimensional condensates, as it is the simplest model that guarantees the modular symmetry of the experimental systems. Here we formulate a %precise 
lattice electric-field representation of the HXY model and contrast this with an analogous representation of the Villain model and the two-dimensional Coulomb gas with a purely rotational auxiliary field.  We find that the HXY model is a %two-dimensional 
spin-model analogue of a lattice electric-field model of the Coulomb gas with an auxiliary field, but with a temperature-dependent vacuum (electric) permittivity that encodes the coupling of the spin vortices to their background spin-wave medium. The spin vortices map to the Coulomb charges, while the spin-wave fluctuations correspond to auxiliary-field fluctuations. The coupling explains the striking differences in the high-temperature asymptotes of the specific heats of the HXY model and the Coulomb gas with an auxiliary field. 
Our results elucidate the propagation of effective long-range interactions throughout the HXY model (whose interactions are purely local) by the lattice electric fields.  They also imply that global spin-twist excitations (topological-sector fluctuations) generated by local spin dynamics are ergodically excluded in the low-temperature phase. We discuss the relevance of these results to condensate physics.  

\end{abstract}

% Uncomment for PACS numbers
%
% Uncomment for keywords
%\vspace{2pc}
%\usepackage{amsmath}
% Uncomment for Submitted to journal title message
\submitto{\JPCM}
%
% Uncomment if a separate title page is required
%\maketitle
% 
% For two-column output uncomment the next line and choose [10pt] rather than [12pt] in the \documentclass declaration
\ioptwocol

\section{Introduction}
\label{intro}
The classical two-dimensional XY (2D-XY) spin model remains an area of active interest in condensed-matter physics. This is fuelled by its remarkable statistical mechanics~\cite{Berezinskii,KTNov,Kosterlitz,JKKN} and its applicability to thin-film and layered ferromagnetic~\cite{BramwellHoldsworth,Andrea}, superfluid~\cite{Nelson_Kosterlitz,BishopReppy,Minnhagen1987Review2DCG} and superconducting~\cite{Beasley,Wolf_superconduct,Resnick,Minnhagen_superconduct_rapidPRB_1981,Minnhagen1987Review2DCG,Rasolt1991KTTransitionAndChargeRedistribution,Baity2016Effective2dThicknessCuprate} phases, along with a wide variety of other experimental systems~\cite{Trombettoni,Hadzibabic,Birgeneau,Sondhi_Bose_review,vinokur2008superinsulator,Baturina2}. %, most notably superconductors and superfluids. 
The system has a divergent spin-spin correlation length (and is therefore critical) in the low-temperature phase but is paramagnetic above the Berezinskii-Kosterlitz-Thouless (BKT) transition~\cite{Berezinskii,KTNov} temperature.  As the system passes through the transition from the low-temperature phase, tightly bound pairs of spin vortices -- the local topological defects of the system -- 
unbind and destroy the quasi-long-range order associated with the divergent correlation length.

Kosterlitz and Thouless (KT) discovered~\cite{KTNov,Kosterlitz} that the 2D-XY model is physically very similar to the two-dimensional Coulomb gas, where the BKT transition becomes the charge confinement-deconfinement transition first discovered by Salzberg and Prager~\cite{Salzberg}. Nelson and Kosterlitz~\cite{Nelson_Kosterlitz} later extended this concept to two-dimensional superfluids and other condensates, in which the spins become the phases of the condensate wavefunction and the spin vortices becomes vortices in the superfluid velocity field. Ambegeokar {\it et al.}~\cite{Ambegaokar1978DissipationInTwo-dimensionalSuperfluids} used the Coulomb-gas %mapping 
description to derive a detailed theory of superfluid linear response, which was used to interpret experiments on helium-4 films~\cite{BishopReppy}. The Hamiltonian of the Coulomb system is, however, quadratic, whereas the classical 2D-XY spins interact via a cosine potential.

In analytical terms, the original BKT ideas were considerably sharpened by  the work of Villain~\cite{Villain}, who constructed a model %with the same 
to approximate the 2D-XY partition function %as the 2D-XY model but 
with a purely quadratic Hamiltonian. Jos\'e {\it et al.}~\cite{JKKN} used the Villain model to decouple the local topological defects from the spin waves (linear phase fluctuations) %and %spin vortices
thereby 
demonstrating %the
a mapping between the 2D-XY model and the two-dimensional Coulomb gas. Vallat and Beck (VB) later established %the precise 
mappings for the Villain and related models on the torus, %(or with periodic boundary conditions) 
and clarified many aspects of the Coulomb-gas description~\cite{VB}. 

%A question that one might reasonably ask is: is this decoupling between the spin waves %(linear phase fluctuations) 
%and spin vortices really valid, and what would be the consequences of it not being valid? Clearly it is not valid at a microscopic level, because each spin difference is modular on $2 \pi$, so that the spin vortices and spin waves can never be truly statistically independent in a microscopic spin model. The Villain model sidesteps this issue by not being a microscopic spin model -- it is better understood as a transformation of the partition function. We also note that, while having the property of modular periodicity, it does not necessarily retain the modular symmetry of a microscopic spin or phase model.  

A question that one might reasonably ask is: is this decoupling between the spin vortices and spin waves really valid, and what would be the consequences of it not being valid? Clearly it is not valid at a microscopic level, because the internal energy of each spin difference is $2\pi$-modular symmetric, so that the spin vortices and spin waves can never be truly statistically independent in a microscopic spin model constructed purely from a set of spins. The Villain model sidesteps this issue by introducing a set of integer-valued variables that allow the Villain analogue of each spin difference to explore all values along the real line. The price that is paid for this is that the modular symmetry on the spin differences %of an XY spin model constructed purely from a system of spins 
is relegated to the more general property of modular periodicity. 
%\blue{A generalized electrostatic model has modular periodicity if each electric-flux component is able to make lone discrete jumps by integer multiples of the elementary charge q; it has modular symmetry if the energy of the electric-flux components are equivalent modulo q. Symmetry implies periodicity but periodicity does not imply symmetry.}

The simplest microscopic spin model that  maintains the modular symmetry of a real spin or condensate system while retaining the Villain model's attractive characteristic of  having a purely quadratic Hamiltonian is the two-dimensional harmonic XY (HXY) model. This is a `pure spin model' in the sense that it is constructed purely from a set of spins.  %in which 
These classical spins interact via a piecewise parabolic potential.  The HXY model was introduced independently by VB~\cite{VB_PRL,VB}, who refer to it as the piecewise parabolic model, and by two of the present authors~\cite{BramwellHoldsworthHXY}. Compared to the Villain model, the HXY model should be seen as the more realistic classical lattice spin model of condensate physics, as well as being more easily visualized and more easily simulated; compared to the 2D-XY model, it avoids the many complications of non-linear spin waves and anharmonic couplings. %, and is thus a worthy object of attention. %Although it is less well known or studied than the Villain model, the HXY model should probably be seen as the canonical classical spin model of two-dimensional condensates, and is thus a worthy object of attention.  
%We therefore 
It is therefore a worthy object of attention and we consider it a valuable exercise to elucidate the relationship between the HXY model and the two-dimensional Coulomb gas, which is the principal aim of this work. 

Our method is a development of that introduced in a previous paper, where we showed %how 
that the global topological properties of the BKT transition in the Coulomb gas on the torus can be understood by extending its phase space to include a freely fluctuating rotational or `auxiliary' field~\cite{FBH}. This allowed us to map the problem on to the generalized lattice electric-field model that was introduced by Maggs and co-workers~\cite{MR,RossettoThesis,Auxiliary,MRottler,LARM,LM} as a computational tool for simulating Coulomb fluids.  An advantage of applying our lattice electric-field method to the present problem is that it allows us to directly connect with a discovery that we made in our previous work: that the ergodic freezing of topological-sector fluctuations is a powerful signature of the BKT transition. Topological sectors are global topological defects in the lattice electric field, which can be generated by charges tracing closed paths around the torus. This idea has subsequently been extended in detail to a model of strongly interacting Bosons in a ring lattice~\cite{Roscilde2016FromQuantum} and has been suggested as a mechanism for the non-ergodic dynamics observed experimentally in layered cuprates at the superconducting transition~\cite{Shi2016EvidenceCorrelatedDynamicsCuprates}. Hence, the lattice electric-field method  allows us to identify the nature of the topological-sector fluctuations of the HXY model, which are of direct relevance to fluctuations in experimental spin systems and condensates. A similar lattice-field model to that of Maggs and co-workers has been applied in two spatial dimensions~\cite{Raghu}.  

There are also more general motivations for the present study. %First, in view of recent interest in the coupling between the phase and amplitude fluctuations in the condensate wavefunction~\cite{Jakubczyk2016AmplitudeFluctuations}, it should be useful to understand the properties of the pure phase (or spin) model that most closely approximates condensate physics on a lattice. In the future, this could be used as a benchmark for contrasting the role of amplitude and phase fluctuations in such systems. 
%Second, we remark that t
The generalized lattice electric (or electromagnetic) field description of spin systems in fact has %rather general 
wider relevance. %-- f
For example, it applies to the description of spin ice and other frustrated magnets in three spatial dimensions~\cite{Ryzhkin,CMS,BrooksBartlett}. Hence, it seems relevant  to showcase the power and simplicity of such a description in one definite case: the two-dimensional HXY model. %in two spatial dimensions.

Our main conclusions may be summarized as follows. While the Villain model with a temperature-independent coupling constant maps exactly to the two-dimensional Coulomb gas with a freely fluctuating auxiliary field, the HXY model, as anticipated, does not. 
Such an exact mapping is easily seen to be impossible in a pure spin model, since the specific heat of a spin model with bounded spin differences is necessarily zero in the high-temperature limit, while a freely fluctuating auxiliary field contributes a constant term due to equipartition of energy. However, the HXY model is found to be a two-dimensional spin-model analogue of the Maggs-Rossetto (MR) generalized lattice electric-field model of Coulomb fluids~\cite{MR}, with a temperature-dependent vacuum (electric) permittivity that encodes the coupling between the HXY spin vortices and the background spin-wave medium. This permittivity could be a real object in experimental condensates. Our lattice electric-field representation further reveals that topological-sector fluctuations in the two-dimensional Coulomb gas with a freely fluctuating auxiliary field~\cite{FBH} correspond to global HXY spin-twist excitations, from which it follows that global HXY spin-twist excitations generated by local spin dynamics are ergodically excluded in the low-temperature phase. 

%\red{May need to improve the following passage.}
%Our main conclusions may be summarised as follows. While the constant-J Villain model maps exactly to the two-dimensional Coulomb gas with a freely fluctuating auxiliary gauge field, the HXY model, as anticipated, does not. %-- indeed s
%Such an exact  mapping is easily seen to be impossible in a pure spin model\blue{, since} the specific heat of a spin model with bounded spin differences is necessarily zero in the high-temperature limit, while a freely fluctuating auxiliary gauge field contributes a constant term due to equipartition of energy. However, the harmonic mode of the HXY lattice electric field is found to be essentially Coulombically correlated in the low-temperature phase, with a temperature-dependent vacuum (electric) permittivity that encodes the coupling of the two-dimensional Coulomb gas of elementary spin vortices to its background spin-wave medium. This permittivity could be a real object in experimental condensates. Our lattice electric-field representation further reveals that topological-sector fluctuations in the standard two-dimensional Coulomb gas~\cite{FBH} correspond to global HXY spin-twist excitations, from which it follows that global HXY spin-twist excitations %, when  generated by local spin dynamics, 
%generated by local spin dynamics are ergodically excluded in the low-temperature `Coulomb' phase. 

The paper is organized as follows. In Section \ref{background}, we review the two-dimensional grand-canonical MR electrostatic model and discuss its equivalence to the Villain model with a temperature-independent coupling constant.  In Section \ref{sec:IntroHXY}, we introduce the HXY model and discuss its excitations. In Section \ref{sec:Coulomb}, we define the MR-type lattice electric-field representation of the HXY model, before showing that its Coulomb and auxiliary-field (spin-wave) components energetically decouple from one another. We then compute its helicity modulus and present numerical data that exhibits the BKT universal jump. 
%, before defining the temperature-dependent HXY vacuum permittivity and performing an analysis to show that the HXY model can be 
%\blue{In Section \ref{sec:CoulombPhase}
%Following this, the harmonic mode of the lattice electric field is shown to be approximately Coulombically correlated in the low-temperature Coulomb phase, but with a temperature-dependent vacuum permittivity that encodes the coupling of the spin vortices to the background spin-wave medium. 
Following this, the HXY model is shown to be a spin-model analogue of the MR model, but with a temperature-dependent vacuum permittivity that encodes the coupling of the spin vortices to the background spin-wave medium. We then present numerical data of the specific heats of the HXY and MR models and explain how the coupling between the HXY spin vortices and spin waves leads to the qualitative differences in their high-temperature asymptotes, before showing that HXY topological-sector fluctuations have the same qualitative behaviour as those of the %precise 
MR  model. In Section \ref{twist}, we present an analysis of the microscopic mechanics of the HXY model, from which we elucidate the propagation of long-range interactions throughout the system before inferring that global HXY spin-twist excitations generated by local spin dynamics are ergodically %frozen out
excluded in the low-temperature phase. %of the BKT transition. 
%topological-sector fluctuations in the lattice electric-field representation of both models correspond to global twist fluctuations in their spin-field representations. 
Conclusions and comparisons with experimental systems are discussed in Section \ref{conclusions}.

Throughout this paper, %all functions are defined to be the discrete counterparts of smooth vector fields~\cite{Chew}, and 
all lattice vector fields %$\mathbf{F}$ is defined~\cite{Chew} component-wise via
%\begin{eqnarray}
%\mathbf{F}(\mathbf{x}):=F_x\left( \mathbf{x}+\frac{a}{2}\mathbf{e}_x\right) \, \mathbf{e}_x + F_y\left( \mathbf{x}+\frac{a}{2}\mathbf{e}_y\right) \, \mathbf{e}_y,
%\end{eqnarray}
%where $\mathbf{x}$ is any lattice site and $\mathbf{e}_{x/y}$ is the unit vector in the $x/y$ direction. 
are defined using the formalism of Chew~\cite{Chew}. 
As a consequence of this definition, the lattice vector-field components exist at the centres of the bonds that connect the lattice sites. 
%The Lesbegue measure $\int \mathcal{D}\mathbf{F}$ of any lattice vector field $\mathbf{F}$ whose components may take values along the whole real line at each lattice site is then %given by
%\begin{eqnarray}
%\int \! \mathcal{D}\mathbf{F}\! :=\! \prod_{\mathbf{x}\in D} \left[ \int_{\mathbb{R}}dF_x(\mathbf{x}\!+\! a\mathbf{e}_x/2)\! \int_{\mathbb{R}} dF_y(\mathbf{x}\!+\! a\mathbf{e}_y/2) \right]\! .
%\end{eqnarray}
%defined as in ref.~\cite{FBH}. 
The operators $\boldsymbol{\tilde{\nabla}}$ and $\boldsymbol{\hat{\nabla}}$ are the forwards and backwards finite-difference operators on a lattice, respectively, and the lattice Laplacian is defined by $\boldsymbol{\nabla}^2:=\boldsymbol{\hat{\nabla}} \cdot \boldsymbol{\tilde{\nabla}}$~\cite{Chew}. All quantities that map between the MR and HXY systems will be assigned the same notation. All lattices are two-dimensional squares with periodic boundary conditions (PBCs) applied. The PBCs enforce the toroidal topology but the curvature of a true torus is not considered. 
The total charge of each system considered here is zero. For the MR models, we work in the grand-canonical ensemble.

\section{The Maggs-Rossetto and Villain models}
\label{background}

%\subsection{The Maggs-Rossetto electrostatic model}\label{MR_background}

The Maggs-Rossetto (MR) generalized electrostatic model~\cite{MR} %-- developed by Maggs and co-workers~\cite{MR,RossettoThesis,Auxiliary,MRottler,LARM,LM} -- 
transforms the lattice Coulomb fluid into a local problem %via the introduction of 
by introducing 
a freely fluctuating, divergence-free auxiliary field to the system. %, whose purpose is outlined in the final paragraph of this section. 
The total lattice electric field $\mathbf{E}$ is Helmholtz decomposed into its Poisson (divergence-full) $-\boldsymbol{\tilde{\nabla}}\phi$, harmonic $\mathbf{\bar{E}}$, and auxiliary-field (rotational) $\mathbf{\tilde{E}}$ components. This total electric field is the most general solution to Gauss' law of lattice electrostatics. The sum of the Poisson and harmonic components of the total electric field is then the electrostatic (or Coulomb) solution to Gauss' law, which describes the Coulomb charges and will be referred to as the Coulomb component of the lattice electric field throughout.

%\begin{eqnarray}\label{MR_Helmholtz}
%\mathbf{E}(\mathbf{x}) = -\boldsymbol{\tilde{\nabla}}\phi(\mathbf{x})+\mathbf{\bar{E}}+\mathbf{\tilde{E}}(\mathbf{x}).
%\end{eqnarray}
%In two spatial dimensions, this lattice electric field is the most general solution to Gauss' law of two-dimensional lattice electrostatics:
%\begin{eqnarray}
%\label{gauss}
%\boldsymbol{\hat{\nabla}} \cdot \mathbf{E}(\mathbf{x})=\rho(\mathbf{x})/ \epsilon_0,
%\end{eqnarray}
%where $\rho (\mathbf{x}):= q m(\mathbf{x})/a^2$ is the charge density, $q$ is the elementary charge, $m(\mathbf{x})\in \mathbb{Z}$ is the charge value at $\mathbf{x}$ in units of $q$, $a$ is the lattice spacing, and $\epsilon_0$ is the electric permittivity of the vacuum~\cite{FBH}. In the above, 

The grand partition function %given by eq. (\ref{grandPF}) 
of the MR model is separable~\cite{MR,FBH} into its Coulomb and auxiliary-field components so that the charge correlations are statistically independent of the auxiliary field $\mathbf{\tilde{E}}$. %: this allows the auxiliary field to be introduced to the model. 
The resultant MR algorithm~\cite{MR} then consists of two types of local lattice electric-field update. The first type of update is the charge-hop update: for a positive elementary electric charge hopping from charge site $\alpha$ to an adjacent charge site $\beta$, the corresponding lattice electric-field update is given by $E_{\alpha \beta}\mapsto E_{\alpha \beta}-q/ \epsilon_0$, where $E_{\alpha \beta}$ denotes the %lattice 
electric %-field 
{\it flux} flowing directly from site $\alpha$ to site $\beta$ without passing through any other charge sites, $q$ is the elementary charge and $\epsilon_0$ is the electric permittivity of the vacuum. 
A generalized electrostatic model has modular periodicity if each lattice electric-field component is able to make lone discrete jumps by integer multiples of some constant; it has modular symmetry if the internal energy of each lattice electric-field component is equivalent modulo this constant. Modular symmetry implies modular periodicity, but modular periodicity does not imply modular symmetry. 

The charge-hop update alters %all degrees of freedom of the 
both the Coulomb and auxiliary-field components of the electric field, thereby producing a greater change in the total internal energy density than %the energy change due to %electrostatic-field updates 
that due to a change in the Coulomb component alone. The algorithm therefore separately samples the auxiliary field in order to allow the total fields to relax to field configurations of lower internal energy. This involves randomly selecting a lattice plaquette and then proposing a rotation of the lattice electric field around the plaquette such that the new field configuration still satisfies Gauss' law for the unchanged charge configuration. These two local updates combine to %sufficiently 
explore the extended phase space of the Coulomb %problem
system and auxiliary field, thereby propagating the long-range Coulomb interactions throughout the system~\cite{MR}. %\red{Is this OK? MR explores the phase space of the Coulomb problem plus small phase-space fluctuations around the problem, corresponding to aux.-field fluctuations. ALSO, we could restrict to solely discussing models and fluctuations, rather than algorithms and updates - could give Tony more credit, and is more general?}

The harmonic mode of the lattice electric field $\mathbf{\bar{E}}:=\sum_{\mathbf{x}\in D}\mathbf{E}(\mathbf{x})/N$ introduced above is the mean lattice electric field averaged over the whole lattice, where $D$ and $N$ are the set and number of charge-lattice sites, respectively. For the case of the two-dimensional MR model of elementary charges, one may adopt the origin-independent model~\cite{VB,FBH} of the total harmonic mode of the lattice electric field, which 
is given by the sum of two components:
%\begin{eqnarray}\label{Ebar_pol_relationCG}
%\mathbf{\bar{E}}=-\frac{1}{\epsilon_0}\mathbf{P}+\frac{q}{L\epsilon_0}\mathbf{w_0},
%\end{eqnarray}
\begin{eqnarray}\label{eq:origin-independentEbar}
\mathbf{\bar{E}}=\mathbf{\bar{E}}_{\rm p}+\frac{q}{L\epsilon_0}\mathbf{w}.
\end{eqnarray}
%where 
Here, $\mathbf{\bar{E}}_{\rm p}$ is the origin-independent polarization component of the harmonic mode, %$q$ is the elementary charge, 
$L$ is the linear system size %$\epsilon_0$ is the electric permittivity of the vacuum, 
and $\mathbf{w}$ is the origin-independent integer-valued winding field.
%The above 
%This formulation adopts the origin-independent model~\cite{VB,FBH} of the total harmonic mode of the lattice electric field, where $\mathbf{\bar{E}}_{\rm p}$ is the origin-independent polarization component of the harmonic mode and $\mathbf{w}$ is the origin-independent integer-valued winding field. 
The winding field $\mathbf{w}$ is fixed by the condition $\bar{E}_{{\rm p},x/y}\in \left( -q/2L\epsilon_0,q/2L\epsilon_0\right]$ and defines the global topological sector of the system, which changes when a charge traces a closed path around the torus. For a given charge configuration, the polarization component $\mathbf{\bar{E}}_{\rm p}$ is then the low-energy solution of the harmonic mode of the lattice electric field. %We note that this origin-independent model may only be adopted for a two-component neutral Coulomb system.

%It follows~\cite{VB} from eq. (\ref{grandPF}) that the inverse electric permittivity of the system $\epsilon^{-1}$ is given by the response function,
%\begin{eqnarray}\label{def_eff_perm}
%%\epsilon_{\rm eff}^{-1}(L,T)=\frac{1}{\epsilon_0^2L^2} \left. \frac{\partial^2 \Phi (L,T,D_i)}{\partial D_i^2}\right|_{D_i\rightarrow 0},
%\epsilon^{-1}(L,T)=\frac{1}{\epsilon_0^2L^2} \left. \frac{\partial^2 \Phi (L,T,D_{\mu})}{\partial D_{\mu}^2}\right|_{D_{\mu}\rightarrow 0},
%\end{eqnarray}
%where $\Phi$ is the grand potential of the system under the influence of a small global applied field $D_{\mu}$ ($\mu \in \{ x,y\}$). This response function measures the response of the harmonic mode of the electric field (due to the charges) to a global applied field in the limit of vanishing applied field. %I

%From eq. (\ref{grandPF}) it can be shown
The harmonic mode encodes all information about the electric permittivity. %, which is a function of the charge correlations. 
It is straightforward to show~\cite{VB} that the inverse electric permittivity of the two-dimensional Coulomb gas %reduces to
is given by
\begin{eqnarray}\label{inverse_eff_perm}
%\epsilon_{\rm eff}^{-1}(L,T) = \epsilon_0^{-1}\left[  1-\frac{1}{2}\chi_{\mathbf{\bar{E}}}(L,T) \right] ,
\epsilon^{-1}(L,T) = \epsilon_0^{-1}\left[  1-\frac{1}{2}\chi_{\mathbf{\bar{E}}}(L,T) \right] ,
\end{eqnarray}
where $\chi_{\mathbf{\bar{E}}}(L,T):=\beta \epsilon_0L^2\left( \langle \mathbf{\bar{E}}^2\rangle - \langle \mathbf{\bar{E}}\rangle^2  \right)$ is the harmonic-mode susceptibility of the lattice electric field. %, and $\langle \dots \rangle_{\rm MR}$ defines the MR thermal average, for which each component of the lattice electric field at each charge site may take any value along the real line (as opposed to an HXY thermal average, for which the lattice electric-field components will be constrained to the set $\left[ -\pi ,\pi \right]$) and each charge site may be occupied by a charge value of any integer multiple of the elementary charge $q$. 
The inverse electric permittivity $1/\epsilon$ measures the response of the harmonic mode of the electric field (due to the charges) to a global applied field in the zero-applied-field limit. %of vanishing applied field. 
In a simply connected space, $\mathbf{\bar{E}}=-\sum_{\mathbf{x}\in D}\mathbf{x}\rho(\mathbf{x})/N\epsilon_0$, from which it follows that the electric permittivity is intimately related to the charge correlations, and is therefore a signature of Coulomb physics. In the above, $\rho (\mathbf{x}):= q m(\mathbf{x})/a^2$ is the charge density at $\bf x$ and $m({\bf x})\in \mathbb{Z}$ is the charge value at $\bf x$ in units of the elementary charge $q$. It follows that $\boldsymbol{\hat{\nabla}}\cdot \mathbf{E}(\mathbf{x})=\rho(\mathbf{x})/\epsilon_0$, i.e., Gauss' law on the lattice is confirmed. 

The Villain model is defined by the short-ranged interaction potential given by eq. (2.4) of ref.~\cite{Villain}. 
We now establish the relationship between the MR and Villain models. Sets of microscopic variables were introduced~\cite{FBH} to represent the two local updates of the algorithm proposed by MR (outlined above). A conjugate lattice $D'$ is defined such that each of its sites is at the centre of a plaquette in $D$ (with a one-to-one correspondence). Each site in $D'$ is associated with a real-valued variable $\varphi \in (-q/2,\, q/2]$ whose adjustment corresponds to an auxiliary-field rotation around the plaquette in $D$ on which the site in $D'$ exists, while each pair of nearest-neighbour sites in $D'$ is associated with an integer-valued variable $s$ whose adjustment corresponds to a charge-hop update in the MR algorithm (this is depicted in figs. 5 and 6 of ref.~\cite{FBH}). %\blue{When $s(\mathbf{x},\mathbf{x}')=0\, \forall \mathbf{x},\mathbf{x}'\in D'$, zero charge exists in the system; any modular $s$ variable changing from zero to one would then correspond to the excitation of a neutral charge pair from the vacuum.}  
Both sets of variables are subject to PBCs. In this microscopic-variable representation, the grand partition function of the two-dimensional, grand-canonical MR  model of multi-valued Coulomb charges is given by~\cite{FBH} 
\begin{eqnarray}\label{MR_microPF}
Z\!=\! \sum_{\{ s\} }\!\int \! \mathcal{D}&\varphi& \!\! \exp \left[ \! - \frac{\beta}{2\epsilon_0}\!\sum_{\langle \mathbf{x},\mathbf{x}'\rangle}\! |\varphi(\mathbf{x})\!-\!\varphi(\mathbf{x}')\!+\!qs(\mathbf{x},\mathbf{x}')|^2 \! \right] \nonumber\\
&\times& \!\! \exp \left( -\beta U_{\text{Core}}\right),
\end{eqnarray}
where $\sum_{\langle \mathbf{x},\mathbf{x}'\rangle }$ is the sum over all nearest-neighbour lattice sites $\mathbf{x},\mathbf{x'} \in D'$, the sets $\{ s(\mathbf{x},\mathbf{x}') \}$ and $\{ \varphi (\mathbf{x})\}$ are the sets of microscopic variables outlined above, $\sum_{\{ s\} }:=\sum_{\{ s(\mathbf{x},\mathbf{x}')\in \mathbb{Z}\} }$ is the sum over all possible $s$-variable configurations,
\begin{eqnarray}
\int \mathcal{D}\varphi := \prod_{\mathbf{x}\in D'} \left[ \int_{-q/2}^{q/2}d\varphi(\mathbf{x}) \right]
\end{eqnarray}
is the functional integral over all possible $\varphi$-variable configurations, $U_{\rm Core}:=a^4\sum_{{\bf x}\in D}\epsilon_{\rm c}\left[ m({\bf x})\right]\left[ \rho(\mathbf{x})\right]^2/2$ is the core-energy component of the grand-canonical energy %, $n_m$ is the number of charges $mq$, 
and $\epsilon_{\rm c}(m)$ is the core-energy constant for charges $mq$ [with $\epsilon_{\rm c}(m)=\epsilon_{\rm c}(-m)$]. The core-energy component of the grand-canonical energy provides control over the number of each charge species~\cite{FBH}. 
We set $\{  \epsilon_{\rm c}(m)=0 \forall m\in \mathbb{Z}\}$/$\{ \epsilon_{\rm c}(m=0,\pm 1)=0$; $\epsilon_{\rm c}(m\ne 0,\pm 1)=\infty \}$ for the MR model of multi-valued/elementary charges throughout. 
%\blue{When $s(\mathbf{x},\mathbf{x}')=0\, \forall \mathbf{x},\mathbf{x}'\in D'$, zero charge exists in the system; any modular $s$ variable changing from zero to one would then correspond to the excitation of a neutral charge pair from the vacuum.} 
We reiterate that the microscopic variables mimic the local updates of the MR algorithm, and note that the construction of the discrete sum and the functional integral ensure that no microstates are counted more than once. 
%As depicted in figs. 5 and 6 of ref.~\cite{FBH}, a charge hop in the positive $x/y$ direction then corresponds to a decrease/increase in the relevant $s$ variable by an amount $q$,  and an auxiliary-field update corresponds to randomly choosing a site in $D'$, and then altering the $\varphi$ variable at the site by an amount $\Delta$. The latter update rotates the electric flux by an amount $\Delta /\epsilon_0$ around the plaquette. %, leaving Gauss' law satisfied.

Eq. (\ref{MR_microPF}) with $q=2\pi$, $\epsilon_0=1/J$ and $\epsilon_{\rm c}(m)=0 \, \forall m\in \mathbb{Z}$ is the partition function for the short-ranged Villain interaction potential defined by eq. (2.4) of ref.~\cite{Villain} with a temperature-independent interaction coefficient (up to a multiplicative constant). We use $J$ to denote this interaction coefficient and refer to it as the temperature-independent coupling constant.  
This establishes the {\it precise equivalence} between the two-dimensional grand-canonical MR model of multi-valued charges $2\pi m$ %[with its core-energy constants $\epsilon_{\rm c}(m)=0 \, \forall m\in \mathbb{Z}$] 
and the Villain model~\cite{Villain} with a temperature-independent coupling constant $J=1/\epsilon_0$. The Villain spin/integer-valued variables are then the $\varphi$/$s$ variables in eq. (\ref{MR_microPF}).

By comparison with the MR model, it is seen that the three excitations of the Villain model (with a temperature-independent coupling constant) are charges, linear phase fluctuations and topological-sector %(winding field) 
fluctuations (described in Section \ref{intro}).  However, note that the charges (local topological defects) are not spin vortices. This is because spin vortices are defined as local topological defects in a lattice vector field ($\boldsymbol{\Delta}\boldsymbol{\varphi}$; defined below) that is a function of the spins and nothing else. In turn, this is a direct consequence of its system possessing a modular symmetry with respect to each absolute spin difference.  
%$\delta \varphi_{\nu}^{\mathbf{x}}$. 
The two-dimensional MR model in fact only possesses the more general property of modular periodicity, rather than the stronger property of a modular symmetry. A lattice Coulomb gas statistically decoupled from its auxiliary field -- and hence the Villain model with a temperature-independent coupling constant -- does not reflect two-dimensional condensate physics on a microscopic level, where the local topological defects are phase (or spin) vortices due to the modular symmetry of the physical system. In the next section we turn to the HXY model, in which this difficulty is eliminated. 

%\subsection{XY-type models}\label{sec:XYintro}
%\section{The harmonic  XY model and its electric-field representation}\label{sec:EFRepHXY}
\section{The HXY model}\label{sec:IntroHXY}

%Henceforth, in order to avoid bulky notation, all quantities that will be shown to map on to their Coulombic counterparts are assigned the same notation.

%The 2D-XY model of magnetism models the physics of magnetic films. The spin field $\mathbf{s}$, defined by
%\begin{eqnarray}
%\mathbf{s}(\mathbf{x}):=(\cos(\varphi(\mathbf{x})),\, \sin(\varphi(\mathbf{x}))),
%\end{align}
%describes the state of each spin of the 2D-XY model, where $\varphi$ is now the phase of each spin and is referred to as the spin at each lattice point. %(Rather than defining new functions, we redefine those of the previous section in order to avoid bulky notation.) 
%The Hamiltonian of the system $H_{\mathrm{XY}}^0$ is given by minus the sum over the dot products of the spins of each nearest-neighbour pair multiplied by the exchange coupling $J$:
%The 2D-XY model is a magnetic spin model comprised of an ordered array of spins fixed at lattice sites in $D'$, which is now the set of all spin-lattice sites. The Hamiltonian of the model is given by

%\subsection{The model}\label{sec:IntroHXY}

The HXY model is a %magnetic 
spin model comprised of an ordered array of spins fixed at the sites of a square lattice. %$D'$ with constant lattice spacing $a$. %In this section, we show that the model behaves as an effective two-dimensional Coulomb gas of elementary spin vortices \blue{coupled to its background spin-wave medium,} with a temperature-dependent vacuum (electric) permittivity \blue{that encodes this coupling}. 
%In terms of %macroscopic-
%condensate physics, this effective Coulomb system becomes a system of elementary condensate-phase vortices~\cite{Nelson_Kosterlitz} [Gross-Pitaevskii]. %, but we will refer to `spins' rather than `condensate phases' throughout. 
Its Hamiltonian is given by
\begin{eqnarray}\label{eq:HXYHamiltonian}
H_{\mathrm{HXY}}(L)=\frac{J}{2} \sum_{ \mathbf{x}\in D'} |\boldsymbol{\Delta}\boldsymbol{\varphi}(\mathbf{x})|^2,
\end{eqnarray}
where $J>0$ is now the HXY coupling constant, $D'$ is the set of all spin-lattice sites, and $\boldsymbol{\Delta}\boldsymbol{\varphi}$ is the modular spin-difference field, defined (component-wise) via
\begin{eqnarray}\label{eq:DefModSpinDiff}
\left[ \boldsymbol{\Delta \varphi} \right]_{\nu}\! \left( {\bf x}\!+\!\frac{a}{2}\mathbf{e}_{\nu}\! \right)  :=   \left\{
\begin{array}{ll}
    \!\!\!  \delta \varphi_{\nu}^{\mathbf{x}} +2\pi ,& \!\!\! \delta \varphi_{\nu}^{\mathbf{x}}  \in  \left[ -2\pi, -\pi \right]  \!,  \\
     \!\!\! \delta \varphi_{\nu}^{\mathbf{x}} ,& \!\!\! \delta \varphi_{\nu}^{\mathbf{x}}   \in  \left( -\pi, \pi \right]  \!,  \\
     \!\!\!   \delta \varphi_{\nu}^{\mathbf{x}} -2\pi ,& \!\!\! \delta \varphi_{\nu}^{\mathbf{x}}  \in  \left( \pi, 2\pi \right]  \!.  \\
\end{array} 
\right.
\end{eqnarray}
Here, $\delta \varphi_{\nu}^{\mathbf{x}}:= \varphi({\bf x}+a\mathbf{e}_{\nu})-\varphi({\bf x})$ ($\nu \in \{ x,y \}$) is the absolute spin difference %\footnote{The modular operation is applied outside of the interval $\left[ -\pi, \pi \right]$, rather than $\left( -\pi, \pi \right]$. This avoids a bias towards $\left[ \boldsymbol{\Delta}\boldsymbol{\varphi}\right]_{x/y}=\pi$, which would affect the statistics of the helicity modulus of the system.}, 
 and $\varphi \in \left( -\pi ,\pi \right]$ is now the phase of each constituent spin of the system, which we refer to as the spin at each lattice site. %The (modular) spin-difference field $\boldsymbol{\Delta}\boldsymbol{\varphi}$ enforces the modular symmetry required of a pure XY spin model. 
%We highlight that $J>0$ for the ferromagnetic case considered throughout. %\red{N.B. We've changed to $\delta \varphi_{\nu}^{\mathbf{x}}  \! \in \! \left[ -\pi, \pi \right]$ rather than excluding $-\pi$. This allows us to properly account for the fact that $\lim_{x\rightarrow \pm \pi}[V'(x)]=\pm J\pi$ (in VB notation) so that we can use $\langle \mathbf{\bar{E}}^2\rangle$ in $\Upsilon$.} 
Two-dimensional condensates can be modelled with a continuum Hamiltonian that is quadratic in the gradient of the phases of the condensate wavefunction. This assumes that the condensate consists of a large number of Bosons so that amplitude fluctuations in the condensate wavefunction are negligible~\cite{Nelson_Kosterlitz}.
The square-lattice analogue is then the HXY model, whose spins correspond to the condensate phases. 
%Corrections to the assumption of negligible amplitude fluctuations may result in a marginal coupling between the amplitude and phase fluctuations~\cite{Jakubczyk2016AmplitudeFluctuations}.

The above HXY Hamiltonian is composed of two symmetries: one is the global $U(1)$ symmetry, while the other is the $2\pi$-modular symmetry with respect to each absolute spin difference $\delta \varphi_{\nu}^{\mathbf{x}}$. %All three XY-type models considered in this paper have both a $2\pi$-modular periodicity between neighbouring spin sites, along with the global $U(1)$ symmetry. %with respect to each absolute spin difference. 
The modular symmetry allows the system to admit spin vortices, which are local topological defects in the modular spin-difference field $\boldsymbol{\Delta \varphi}$. This local excitation is supplemented with global `spin-twist' excitations %of the \blue{spin} field 
 (topological-sector fluctuations; discussed further in Section \ref{twist}) when the system is placed on the torus. Spin waves are thermal fluctuations around the local minimum-energy field configuration that describes each spin-vortex and topological-sector configuration. %(superposed with the global excitations). 
%These two local excitations are supplemented with global \blue{`twist'} excitations of the \blue{spin} field on the torus (topological-sector fluctuations), \blue{which are} discussed further in in Section \ref{twist}). 
As noted above, the MR and Villain models also admit three excitations, which are closely related %, but not identical, 
to those of the HXY model. The principal difference is that in the HXY model the local topological defects are spin vortices, whereas in the MR and Villain models they cannot be interpreted in this way (see above). Similarly, the HXY topological sectors (or global topological defects) are global spin-twist excitations, but this is not the case for the MR and Villain topological sectors. We discuss global spin twists in detail in Section \ref{twist}.

Finally, the modular spin-difference field $\boldsymbol{\Delta\varphi}$ is Helmholtz decomposed~\cite{KTNov} into its divergence-free and divergence-full components, where the divergence-free component $\boldsymbol{\Delta \hat{\varphi}}$ describes the local minimum-energy configuration of the total modular spin-difference field %$\boldsymbol{\Delta}\theta$ 
of each spin-vortex configuration (superposed with the global excitations), and the divergence-full component (or spin-wave field) $\boldsymbol{\Delta\psi}$ describes thermal fluctuations around those minima:
\begin{eqnarray}\label{eq:Helmholtz_spin}
\boldsymbol{\Delta\varphi} ({\bf x})= \boldsymbol{\Delta\hat{\varphi}} ({\bf x}) + \boldsymbol{\Delta\psi} ({\bf x}).
\end{eqnarray}
%Eq. (\ref{eq:Helmholtz_spin})  will transform to an equation analogous to eq. (\ref{MR_Helmholtz}) in the electric-field representation of Section \ref{sec:EFrep}, where the divergence-free/divergence-full component will become purely irrotational/rotational.
%This will be informative later in the paper.

%\subsection{The electric-field representation and propagation of long-range interactions}\label{sec:EFrep}
\section{Coulomb physics in the HXY model}\label{sec:Coulomb}

\subsection{Electric-field representation and universal jump}\label{EFrep}

We now define the lattice electric field $\mathbf{E}$ of the HXY model:
\begin{eqnarray}\label{emerg_field}
\mathbf{E}(\mathbf{x}):=\frac{J}{a}
\left[ \begin{array}{c}
[ \boldsymbol{\Delta\varphi} ]_y(\mathbf{x}+a\mathbf{e}_x/2)  \\
\\
-[ \boldsymbol{\Delta\varphi} ]_x(\mathbf{x}+a\mathbf{e}_y/2)  \end{array} \right],
\end{eqnarray}
where $a$ is the lattice spacing and $\mathbf{x}\in D$ is a site on the (conjugate) MR lattice. 
%We Helmholtz decompose the lattice electric field $\mathbf{E}$
The lattice electric field $\mathbf{E}$ is Helmholtz decomposed into its Poisson, harmonic and auxiliary-field components. %in the same form as eq. (\ref{MR_Helmholtz}). 
Combining eqs. (\ref{eq:Helmholtz_spin}) and (\ref{emerg_field}), it then follows that the divergence-free/divergence-full field $\boldsymbol{\Delta\hat{\varphi}}$/$\boldsymbol{\Delta\psi}$ becomes purely irrotational/rotational in the lattice electric-field representation:
\begin{eqnarray}\label{eq:top_def-Coulomb}
-\tilde{\boldsymbol{\nabla}}\phi(\mathbf{x})& +\mathbf{\bar{E}}= & \frac{J}{a}
\left[ \begin{array}{c}
[ \boldsymbol{\Delta\hat{\varphi}} ]_y(\mathbf{x}+a\mathbf{e}_x/2)  \\
\\
-[ \boldsymbol{\Delta\hat{\varphi}} ]_x(\mathbf{x}+a\mathbf{e}_y/2)  \end{array} \right]; \\
%\end{eqnarray}
%\begin{eqnarray}
\label{eq:aux-SW}
& \tilde{\mathbf{E}}(\mathbf{x})= & \frac{J}{a}
\left[ \begin{array}{c}
[ \boldsymbol{\Delta\psi} ]_y(\mathbf{x}+a\mathbf{e}_x/2)  \\
\\
-[ \boldsymbol{\Delta\psi} ]_x(\mathbf{x}+a\mathbf{e}_y/2)  \end{array} \right].
\end{eqnarray}
The (irrotational) Coulomb component of the lattice electric field [eq. (\ref{eq:top_def-Coulomb})] describes the spin vortices with the global excitations superposed, while the (rotational) auxiliary-field component [eq. (\ref{eq:aux-SW})] describes the spin waves. For brevity, we will now refer to quantities expressed in the lattice electric-field representation using the terms of their electrical counterparts.

Combining eqs. (\ref{eq:DefModSpinDiff}) and  (\ref{emerg_field}) then produces %a lattice Gauss' law of the same form as eq. (\ref{gauss}), but with $\epsilon_0=1/J$ and $q=2\pi$. %and with $\rho$ the lThese three excitations also occur in both the (full cosine) 2D-XY model [ref. BKT], where anharmonic terms in the cosine interaction are a complicating factor, and in the Villain model [ref. Villain],  where local topological defects are not true spin vortices as their `winding numbers' may take any integer value (rather than simply $\{0,\pm1\}$).ocal topological-defect density of the XY-type model. 
%XY-type models therefore admit a Gauss' law of the same form as Eq. (\ref{gauss}), but with $\epsilon_0=1/J$, and with %:
%\begin{eqnarray}\label{gaussXY}
%\boldsymbol{\hat{\nabla}}\cdot \mathbf{E}(\mathbf{x})= J\rho(\mathbf{x}),
%\end{eqnarray}
%where 
%$\rho (\mathbf{x}):= 2\pi m(\mathbf{x})/a^2$ now defined to be the local topological-defect density, where $m(\mathbf{x})\in \mathbb{Z}$ is now the local topological-defect value at $\mathbf{x}$ in units of $2\pi$. 
%The HXY model therefore admits 
the lattice Gauss law of two-dimensional electrostatics with $\epsilon_0=1/J$ and $q=2\pi$:
\begin{eqnarray}\label{eq:GaussHXY}
\boldsymbol{\hat{\nabla}}\cdot \mathbf{E}(\mathbf{x})= J\rho(\mathbf{x}),
\end{eqnarray}
where $\rho (\mathbf{x}):= 2\pi m(\mathbf{x})/a^2$ is the charge density at $\bf x$, and $m({\bf x})\in \mathbb{Z}$ is the charge value at $\bf x$ in units of $2\pi$. 
%, which is 
%This is analogous to the first constraint of the MR electrostatic model. %(imposed by the first delta function of eq. (\ref{eq:MR_functional_integral}) in the MR case). 
Further to this, the harmonic mode of the lattice electric field is of the same form as eq. (\ref{eq:origin-independentEbar}), but again with $\epsilon_0=1/J$ and $q=2\pi$, where we have made the observation~\cite{VB} that charges given by $m(\mathbf{x})\ne 0,\pm 1$ are not permitted at any charge site in the HXY model. %For a given charge configuration, $\mathbf{\bar{E}}_{\rm p}$ is then the low-energy solution of the harmonic mode of the lattice electric field.

\begin{figure}[t]
\includegraphics[width=\linewidth]{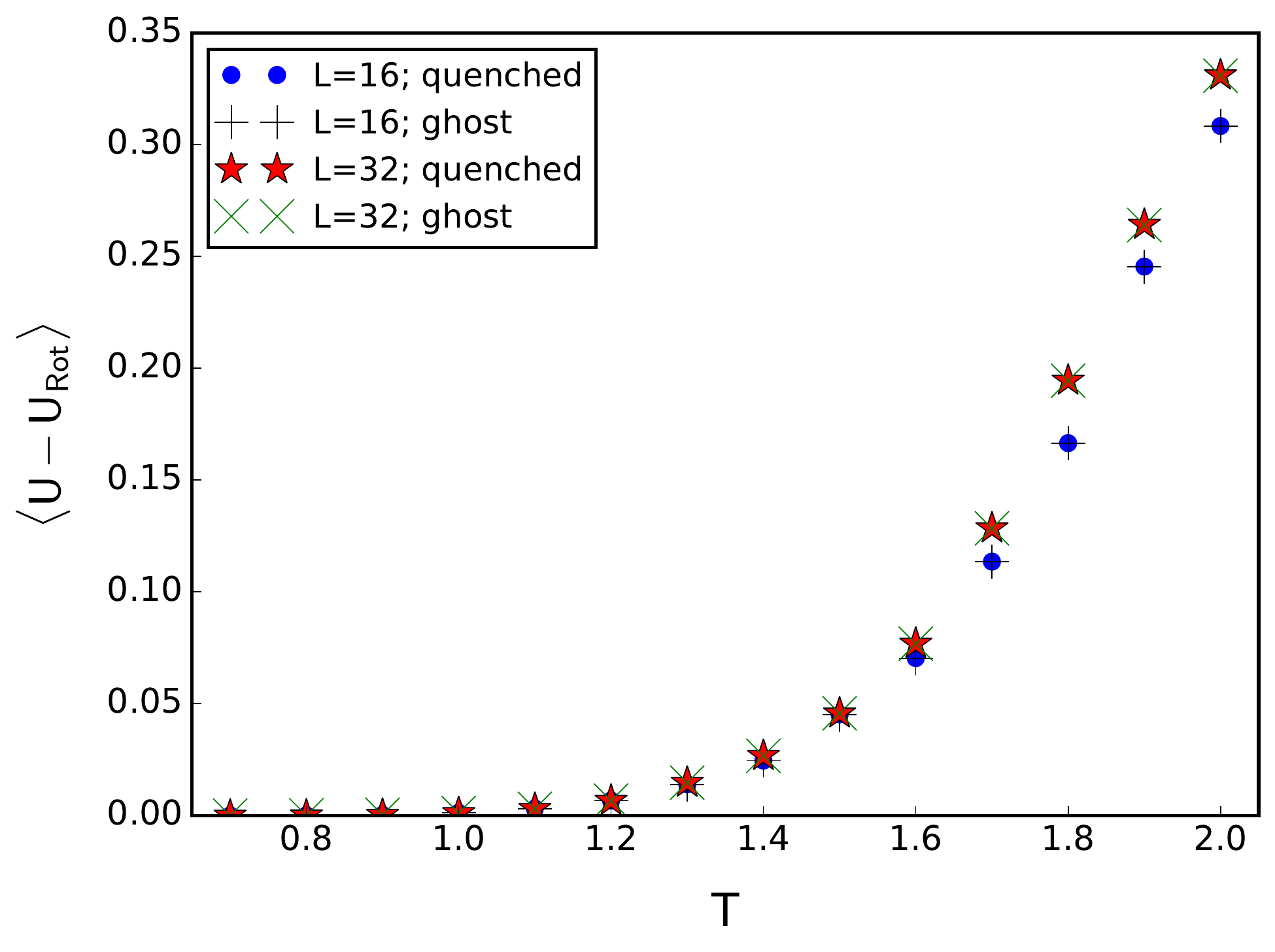}
\caption{Numerical estimates of the internal energy of the Coulomb (irrotational) component of the lattice electric field of HXY systems of linear size $L=16$ and $32$ as functions of temperature $T$ as computed by two different methods. For the first method, we applied a linear solver to the charge configuration of each spin configuration to find the lattice Green's function in eq. (\ref{HXY_emerg_energy1}). %, and 
We  then added the internal energy of the harmonic mode $\mathbf{\bar{E}}$ to the first term in eq. (\ref{HXY_emerg_energy1}) to compute the internal energy without the auxiliary-field term (the `ghost' solution). 
For the second method, we measured the total internal energy after quenching the same spin configuration, where we quenched each spin configuration by performing $10000$ Monte Carlo sweeps at $T=0$.  %(blue stars)  
The data shows  close agreement between the two methods. $U:=H_{\rm HXY}$; $U_{\rm Rot}:=a^2\sum_{\mathbf{x}\in D}|\mathbf{\tilde{E}}(\mathbf{x})|^2/2J$. We do not include error bars as the errors are functions of the simulation method, and here we compare measurement methods. We use units such that $J=k_{\rm B}=a=1$. Simulation details are described in \ref{app:SimDetails}.} \label{fig_energy}
\end{figure}
%We now combine eqs. (\ref{eq:top_def-Coulomb}), (\ref{eq:aux-SW}) and (\ref{eq:GaussHXY}) %with the lattice Gauss' law of the HXY model to write its
Upon combining eqs. (\ref{eq:HXYHamiltonian}) and (\ref{eq:Helmholtz_spin}) with (\ref{eq:top_def-Coulomb})--(\ref{eq:GaussHXY}), the HXY Hamiltonian becomes~\cite{VB}
\begin{eqnarray}\label{HXY_emerg_energy1}
H_{\rm HXY}(L)=&\frac{a^4J}{2}\sum_{\mathbf{x}_i,\mathbf{x}_j\in D}\rho(\mathbf{x}_i)G(\mathbf{x}_i,\mathbf{x}_j)\rho(\mathbf{x}_j) \nonumber\\
&+\frac{L^2}{2J}|\mathbf{\bar{E}}|^2+\frac{a^2}{2J}\sum_{\mathbf{x}\in D}|\mathbf{\tilde{E}}(\mathbf{x})|^2,
\end{eqnarray}
where $G$ is the Green's function of the two-dimensional lattice Coulomb gas. 
Fig. \ref{fig_energy} shows the thermal average of the internal energy of the Coulomb (irrotational) component of the lattice electric field of HXY systems of linear size $L=16$ and $32$ as functions of temperature $T$ as computed by two different methods. For the first method, we applied a linear solver to the charge configuration of each spin configuration to find the lattice Green's function in eq. (\ref{HXY_emerg_energy1}). %, and 
We  then added the internal energy of the harmonic mode $\mathbf{\bar{E}}$ to the first term in eq. (\ref{HXY_emerg_energy1}) to compute the internal energy without the auxiliary-field term. 
For the second method, we measured the total internal energy after quenching the same spin configuration, where we quenched each spin configuration by performing $10000$ Monte Carlo sweeps at $T=0$.  %(blue stars)  
Fig. \ref{fig_energy} shows close agreement between the two methods, thus confirming the energetic decoupling of the HXY lattice electric field into its Coulomb and auxiliary-field components, as in eq. (\ref{HXY_emerg_energy1}). %Quenching an XY-type system removes its spin waves, hence figs. 
%the same Helmholtz-decomposition identifications (eqs. (\ref{eq:aux-SW}) and (\ref{eq:top_def-Coulomb})) follow for the HXY model as for the Villain model.

Using the HXY analogue of the response function that generated eq. (\ref{inverse_eff_perm}), we are able to show that the HXY helicity modulus $\Upsilon$ in the lattice electric-field representation is given by%\footnote{Eq. (\ref{eq:HXYinverse_eff_perm}) is only truly valid in the thermodynamic limit, where a global spin twist corresponding to infinitesimally small spin differences between each pair of adjacent spin sites (in the direction of the global twist) is possible. It is standard practice to apply this expression to the finite-size system before taking the thermodynamic limit.} 
\begin{eqnarray}\label{eq:HXYinverse_eff_perm}
\Upsilon(L,T) = \left[ \epsilon_0^{\rm H} (L,T)\right]^{-1}\left[  1-\frac{1}{2}\chi_{\mathbf{\bar{E}}}^{\rm H}(L,T) \right] ,
\end{eqnarray}
where
\begin{eqnarray}\label{eq:HXYRenormVacuumPerm}
\left[ \epsilon_0^{\rm H} (L,T)\right]^{-1}:= 2J%\sum_{\mathbf{x}\in D}
\sum_{n=1}^{\infty}(-1)&^{n+1}\langle  \cos \left[ \frac{na}{J}E_{x}(\mathbf{0}) \right] \rangle %\right] +\nonumber\\
% & \cos \left[ \frac{na}{J}E_{y}(\mathbf{x})  \right] \rangle
\end{eqnarray}
and
\begin{eqnarray}\label{eq:HXYChiEbar}
\chi_{\mathbf{\bar{E}}}^{\rm H}(L,T):= \beta \epsilon_0^{\rm H}(L,T)L^2\left( \langle \mathbf{\bar{E}}^2\rangle - \langle \mathbf{\bar{E}}\rangle^2  \right)
\end{eqnarray}
are the %temperature-dependent 
temperature-dependent inverse vacuum permittivity and harmonic-mode susceptibility of the HXY model, respectively. The temperature-dependent vacuum permittivity is a function of the second derivative of the Hamiltonian with respect to the each %($x/y$) 
Cartesian component of the lattice electric-field at each lattice site, which are each periodic on $\left(-\pi J/a , \pi J/a \right]$~\cite{VB}. The Fourier series is used to properly account for the piecewise-periodic property of the Hamiltonian, and therefore for the {\it statistical} coupling between the HXY Coulomb and auxiliary fields (described below).
\begin{figure}[t]
\includegraphics[width=\linewidth]{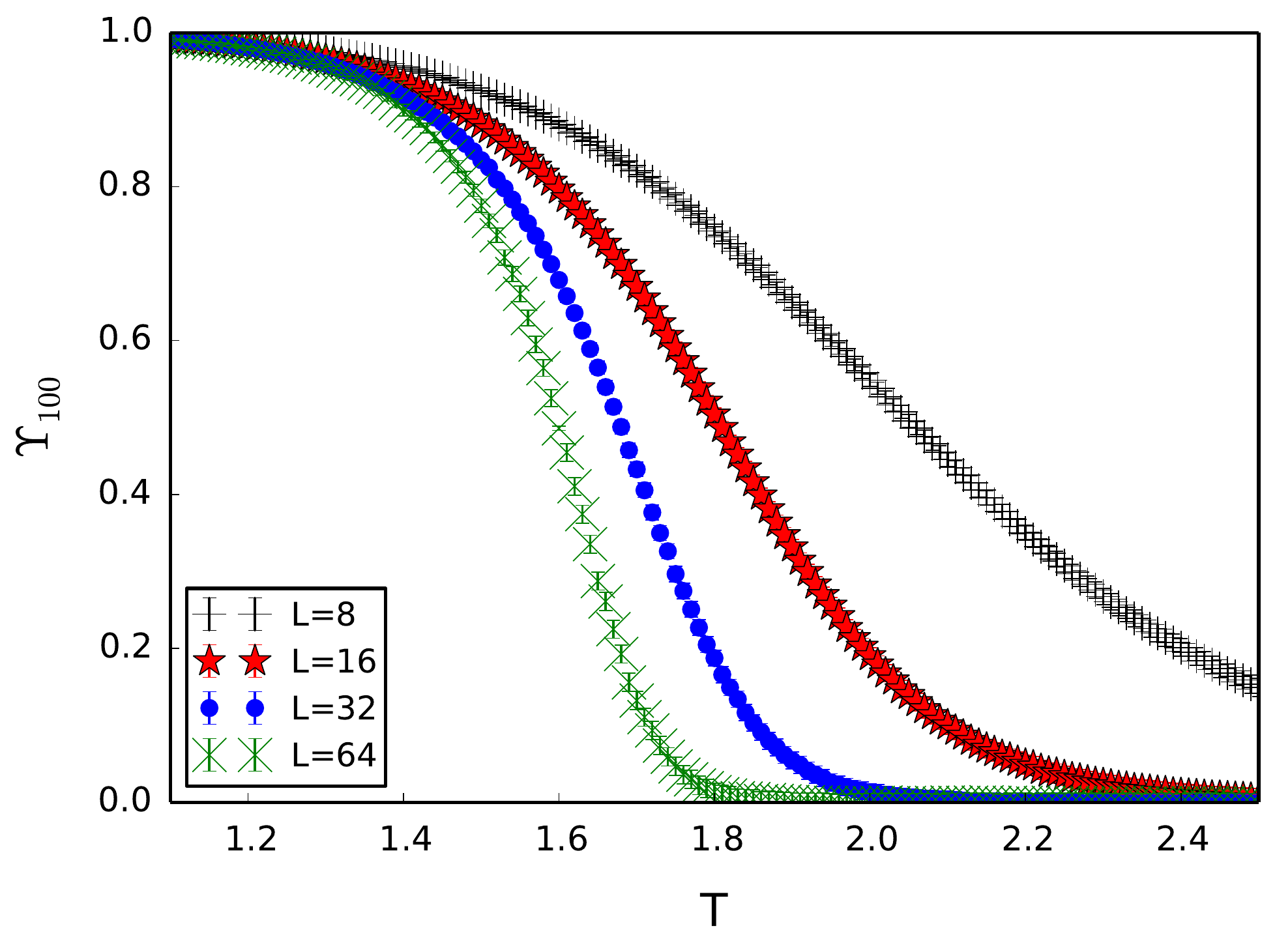}
\caption{Numerical estimates of the HXY helicity modulus $\Upsilon_{100}$ [whose Fourier series [eq. (\ref{eq:HXYRenormVacuumPerm})] has been cut off at $n=100$] %(with the Fourier series of the temperature-dependent HXY vacuum permittivity $1/ \epsilon_0^{\rm H} (L,T)$ cut off at $n=100$) 
 as a function of temperature $T$ for  systems  of linear size $L=8$  to $64$. The data suggests that the helicity modulus $\Upsilon$ tends towards the BKT universal jump at the transition temperature $T_{\rm BKT}\simeq 1.35 $ as system size increases, which is consistent with the expected BKT transition in the HXY model. %confirms that the HXY model does indeed undergo a BKT transition. %The error bars show two standard deviations. 
We use units such that $J=k_{\rm B}=a=1$. Simulation details are described in \ref{app:SimDetails}.}
\label{fig:HXYHelicity}
\end{figure}

Fig. \ref{fig:HXYHelicity} shows the HXY helicity modulus $\Upsilon_{100}$ (whose Fourier series has been cut off at $n=100$) as a function of temperature $T$ for systems  of linear size $L=8$  to $64$. The data suggests that the helicity modulus $\Upsilon$ tends towards the BKT universal jump at the transition temperature $T_{\rm BKT}\simeq 1.35 J/k_{\rm B}$~\cite{Archambault1997MagneticFluctuations} as system size increases, which is consistent with the expected BKT transition in the HXY model. %confirms that the HXY model does indeed undergo a BKT transition.

\subsection{Statistical coupling of the Coulomb and auxiliary fields}\label{sec:CoulombPhase}

In the previous subsection, we confirmed that the Coulomb and auxiliary-field components of the HXY lattice electric field energetically decouple, and also that the system experiences a BKT transition. 
These are both properties of a two-dimensional MR lattice Coulomb gas statistically independent of its auxiliary field. However, the HXY Coulomb and auxiliary-field components are not statistically independent of one another. %This statistical independence is an additional property of a precisely decoupled MR lattice Coulomb gas. %These conditions are necessary but not sufficient for the system to be equivalent to a true two-dimensional Coulomb gas. The Coulomb and auxiliary-field components are in fact not statistically independent of one another, which is a third necessary condition for  Coulomb-gas equivalence. 
This can be seen in the context of a quasi-MR description of the partition function:
\begin{eqnarray}
\label{PartitionHXY}
Z_{\mathrm{HXY}}=&\left(\frac{a}{ J}\right)^{2N}\int \bar{\mathcal{D}}\mathbf{\tilde{e}} \exp \left[ -\frac{\beta a^2}{2J}\sum_{\mathbf{x}\in D}|\mathbf{\tilde{e}}(\mathbf{x})|^2 \right]  \nonumber\\
&\times \!  \exp \left[ -\frac{\beta a^4 J}{2}\sum_{\mathbf{x}_i\ne \mathbf{x}_j}\rho(\mathbf{x}_i)G(\mathbf{x}_i,\mathbf{x}_j)\rho(\mathbf{x}_j) \right]  \nonumber\\
& \times \! \exp\left( -\frac{\beta L^2}{2J}|\mathbf{\bar{E}}_{\rm p}+\frac{2\pi J}{L} \mathbf{w}|^2 \right)  e^{\beta \mu n}  ,
\end{eqnarray}
%in the emergent-field representation, 
which is derived using a similar method~\cite{FBH} to that used to separate the grand partition function of the MR  model into two distinct components\footnote{The $\left( a/ J \right)^{2N}$ factors in the analogues of eq.  (\ref{PartitionHXY}) in the appendices of ref.~\cite{FBH} were missed in error. We also note that all partition functions presented here are functions of linear system size $L$ and temperature $T$.}. Here, $\mu :=-2\pi^2JG({\bf 0})$ is the chemical potential for the introduction of a charge~\cite{FBH}, $G({\bf 0}):=G({\bf x},{\bf x})$ is the diagonal element of the Green's function, $n$ is the number of charges present in the system, and $\int \bar{\mathcal{D}}\mathbf{\tilde{e}}$ is defined via
\begin{eqnarray}
\int \bar{\mathcal{D}}\mathbf{\tilde{e}}:=& \prod_{\mathbf{x}\in D}\int_{-\pi J/a-\hat{E}_x(\mathbf{x}+a\mathbf{e}_{x}/2)}^{\pi J/a-\hat{E}_x(\mathbf{x}+a\mathbf{e}_{x}/2)}d \tilde{e}_x(\mathbf{x}+a\mathbf{e}_x/2) \nonumber\\
&\times \prod_{\mathbf{x}\in D} \int_{-\pi J/a-\hat{E}_y(\mathbf{x}+a\mathbf{e}_{y}/2)}^{\pi J/a-\hat{E}_y(\mathbf{x}+a\mathbf{e}_{y}/2)}d \tilde{e}_y(\mathbf{x}+a\mathbf{e}_y/2 ) \nonumber\\
&\times \sum_{\{ \rho(\mathbf{x})\in \{ 0,\pm 2\pi /a^2  \} \} } \sum_{ \mathbf{w}\in \mathbb{Z}^2 } \delta \left( \sum_{\mathbf{x}\in D}\rho (\mathbf{x})\right) \nonumber\\
&\times \delta \left( \sum_{\mathbf{x}\in D}\mathbf{\tilde{e}}(\mathbf{x})\right)\prod_{\mathbf{x}\in D}\left[ \delta \left(\boldsymbol{\hat{\nabla}}\cdot \mathbf{\tilde{e}}(\mathbf{x})\right) \right] ,
\end{eqnarray}
%\begin{eqnarray}
%\int \bar{\mathcal{D}}\mathbf{\tilde{e}}:=& \int \mathcal{D}\mathbf{\tilde{e}} \sum_{\{ \rho(\mathbf{x})\in \{ 0,\pm 2\pi /a^2  \} \} } \sum_{ \mathbf{w}\in \mathbb{Z}^2 } \delta \left( \sum_{\mathbf{x}\in D}\rho (\mathbf{x})\right) \nonumber\\
%&\times \prod_{\mathbf{x}\in D} \left[ \Theta \left( \pi + \tilde{e}_x (\mathbf{x}+a\mathbf{e}_x/2) + \hat{E}_x (\mathbf{x}+a\mathbf{e}_x/2) \right) \Theta \left( \pi - \tilde{e}_x (\mathbf{x}+a\mathbf{e}_x/2) - \hat{E}_x (\mathbf{x}+a\mathbf{e}_x/2) \right)  \right] \nonumber\\
%&\times \prod_{\mathbf{x}\in D} \left[ \Theta \left( \pi + \tilde{e}_y (\mathbf{x}+a\mathbf{e}_y/2) + \hat{E}_y (\mathbf{x}+a\mathbf{e}_y/2) \right) \Theta \left( \pi - \tilde{e}_y (\mathbf{x}+a\mathbf{e}_y/2) - \hat{E}_y (\mathbf{x}+a\mathbf{e}_y/2) \right)  \right] \nonumber\\
%&\times \delta \left( \sum_{\mathbf{x}\in D}\mathbf{\tilde{e}}(\mathbf{x})\right)\prod_{\mathbf{x}\in D}\left[ \delta \left(\boldsymbol{\hat{\nabla}}\cdot \mathbf{\tilde{e}}(\mathbf{x})\right) \right] ,
%\end{eqnarray}
where $\mathbf{\hat{E}}$ is the Coulomb component of the lattice electric field. %, $\mu := \mu_1$ is the chemical potential for the introduction of an elementary charge, and $n$ is the number of charges. Geometrically, charges given by $m(\mathbf{x})\ne 0,\pm 1$ are not permitted for any charge site: this enforces a core-energy configuration $\{ \epsilon_{\rm c}(m=0,\pm 1)=0,\epsilon_{\rm c}(m\ne 0,\pm 1)=\infty \}$, which corresponds to a system of elementary charges.%(the argument ($\mathbf{x}+a\mathbf{e}_{x/y}/2$) is dropped in the limits of each integral). 

The exponents related to the Coulomb component of the lattice electric field in eq. (\ref{PartitionHXY}) cannot be taken outside of the functional integral over the auxiliary-field component of the lattice electric field: the $\left(-\pi J/a,\, \pi J/a\right]$ bound on both the $x$ and $y$ components of the total lattice electric field at each charge site results in the available configurations of its auxiliary-field component being a function of each charge configuration. %, so that the single measure $\int \bar{\mathcal{D}}\mathbf{\tilde{e}}$ must be applied to the integrand of the HXY partition function. 
Hence, while the Coulomb and auxiliary-field components of the lattice electric field energetically decouple (see fig. \ref{fig_energy}), the partition function itself is not separable into distinct Coulomb and auxiliary-field components. It follows that the Coulomb and auxiliary-field components of the lattice electric field are not statistically independent of one another, and hence that the auxiliary field is non-freely fluctuating. The HXY model is therefore not a precise MR model, but rather a spin-model analogue the MR model with a temperature-dependent vacuum permittivity [eq. (\ref{eq:HXYRenormVacuumPerm})] that encodes the coupling between the HXY Coulomb and auxiliary fields. The coupling is a consequence of the HXY charges being spin vortices, since the $\left(-\pi J/a ,\pi J/a \right]$ bound allows the HXY charges to be defined as local topological defects in a lattice vector field that is a function of the spins and nothing else.

\begin{figure}[t]
\includegraphics[width=\linewidth]{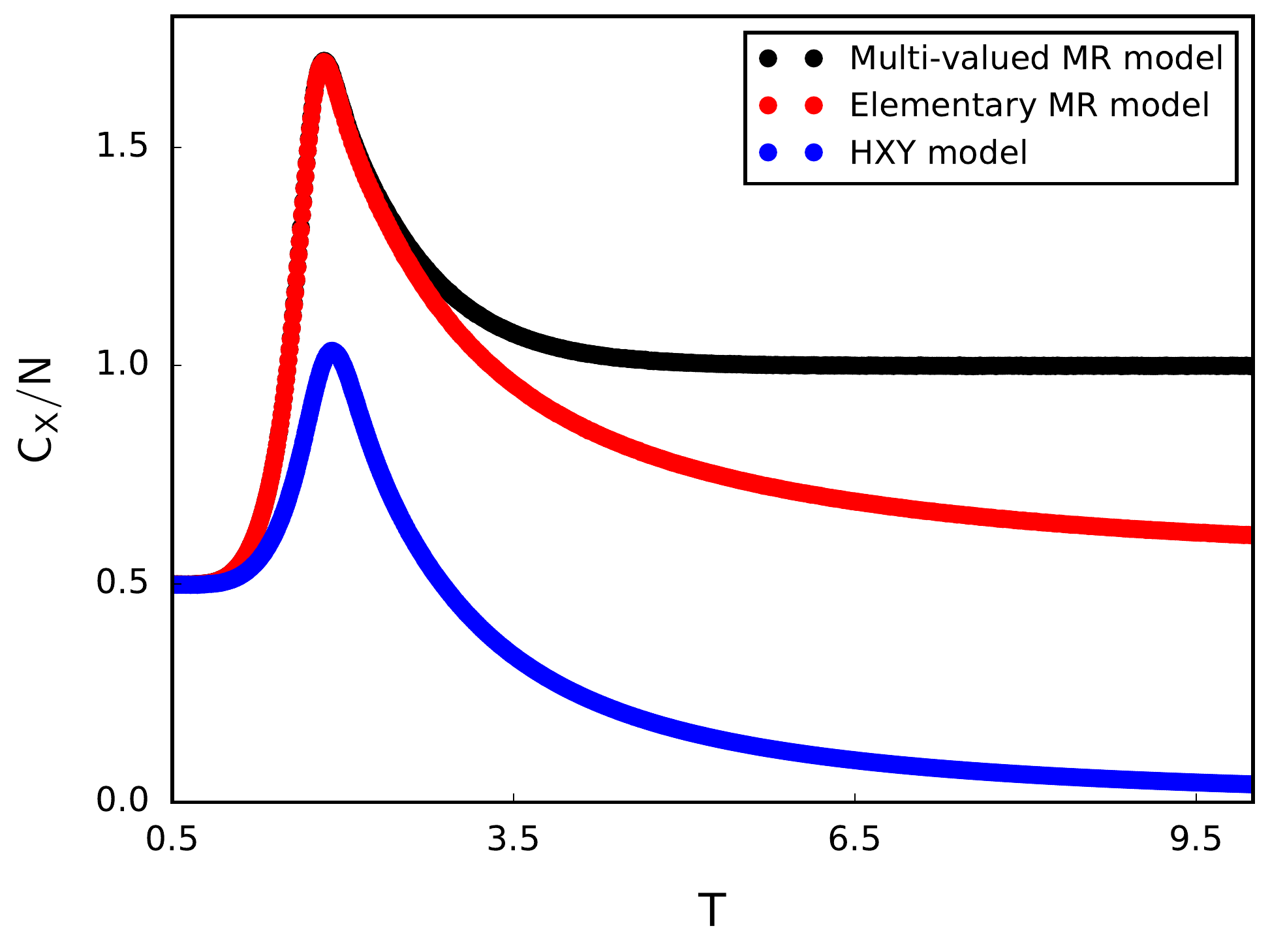}
\caption{Numerical estimates of the specific heats per charge site $C_{\rm X}/N$ of the HXY model and the two-dimensional grand-canonical MR models of both elementary and multi-valued charges as functions of temperature $T$ for systems of linear size $L=16$. $X=H_{\rm ext}$/$\mu_{\rm m}$ for the HXY/MR model, where $H_{\rm ext}$ is the magnitude of an externally applied magnetic field and $\mu_{\rm m}$ is the chemical potential for the introduction of a charge $mq$. The data suggests qualitative differences in the high-temperature asymptotes of each model, which we explain in the text using a mode-counting argument.   We use units such that $J=\epsilon_0=k_{\rm B}=a=1$. Simulation details are described in \ref{app:SimDetails}.} \label{fig:SpecificHeat}
\end{figure}
This has interesting consequences for the specific heat of the HXY model. %\blue{The specific heats $C_{\rm X}$ of the HXY and MR models are both given by 
%\begin{eqnarray}
%C_{\rm X}=\frac{1}{k_{\rm B}T^2}\frac{\partial^2 \ln Z}{\partial \beta^2},
%\end{eqnarray}
%where $Z$ now represents the partition function of each model, and $X=H_{\rm Ext}$/$\mu_{\rm n}$ for the HXY/MR model. Here, $H_{\rm Ext}$ is the magnitude of an externally applied magnetic field and $\mu_{\rm n}$ is the chemical potential for the introduction of a charge $mq$. 
Fig. \ref{fig:SpecificHeat} shows the specific heats $C_{\rm X}$ of the HXY model and the %two-dimensional grand-canonical 
MR models of both elementary and multi-valued charges %\footnote{Elementary MR model: $\epsilon_{\rm c}(m=0,\pm 1)=0$; $\epsilon_{\rm c}(m\ne 0,\pm 1)=\infty$. Multi-valued MR model: $\epsilon_{\rm c}(m)=0 \forall m\in \mathbb{Z}$.} 
as functions of temperature $T$ for systems of linear size $L=16$. Here, $X=H_{\rm ext}$/$\mu_{\rm m}$ for the HXY/MR model, where $H_{\rm ext}$ is the magnitude of an externally applied magnetic field and $\mu_{\rm m}$ is the chemical potential for the introduction of a charge $mq$~\cite{FBH}. The data suggests that the specific heats of the models have equal low-temperature asymptotes [$C_{\rm X}(T \rightarrow 0)=Nk_{\rm B}/2$] but strikingly different high-temperature asymptotes [$C_{\mu_{\rm m}}(T \rightarrow \infty) = Nk_{\rm B}$ (multi-valued charges); $C_{\mu_{\rm m}}(T \rightarrow \infty) = Nk_{\rm B}/2$ (elementary charges); $C_{\rm H_{\rm ext}}(T \rightarrow \infty) = 0$]. %[$C_{\rm \mu_{\rm m}}(T \rightarrow \infty)=Nk_{\rm B}/2$ / $Nk_{\rm B}$ for elementary / multi-valued MR models; $C_{\rm H_{\rm ext}}(T \rightarrow \infty)=0$)]. %, and that e
Each pair of low- and high-temperature asymptotes are separated by a peak. 
%\blue{The data suggests that the high-temperatures asymptotes of the specific heats of the MR model of multi-valued charges, the MR model of elementary charges and the HXY model appears to be $1$, $0.5$ and $0$, respectively, and that the low-temperature asymptotes are all $0.5$.} 
%\blue{The HXY specific-heat data %$C_{H_{\rm ext}}$ %in fact 
%appears to qualitatively resemble the experimental specific-heat data of thin-film liquid helium-4 presented in fig. 43 of %and 1 of refs.
%ref.~\cite{Gasparini2008Finite-sizeScalingHelium}.} %and~\cite{Fisher2010Superfluid}, respectively.}
We now explain these qualitative differences %\blue{between each model} %in the low- and high-temperature limits 
using a mode-counting argument~\cite{Moessner1998PropertiesOfAClassicalSpinLiquid}. 

Each charge site of a hypercubic lattice of an %precise 
MR %electrostatic 
model has $d$ independent lattice electric-field components associated with it, where $d$ is the spatial dimension of the lattice. 
As a consequence of equipartition of energy, each of these $Nd$ Gaussian modes contributes $k_{\rm B}/2$ to the specific heat of the grand-canonical MR model of multi-valued divergences $C_{\mu_{\rm m}}$. In the limit of low temperature, the absence of charge means that $\boldsymbol{\nabla} \cdot \mathbf{E} ({\bf x})=0$ everywhere. This constraint on the lattice electric field at each charge site leaves $N(d-1)$ contributions of $k_{\rm B}/2$ to the specific heat. %so that 
Hence, $C_{\mu_{\rm m}}(T\rightarrow 0) = N(d-1)k_{\rm B}/2 = N k_{\rm B}/2$ for the two-dimensional case. %The $(d-1)$ factor is a direct result of the charge density being constrained to zero everywhere in the limit of low temperature, which corresponds to one constraint per charge site. 
%At high temperature, \blue{the charge density at each lattice site is unbounded, so that} this constraint is lifted and $C_{\rm V}(T \rightarrow \infty) = N k_{\rm B}$. 
At high temperature, if the charge density at each lattice site is unbounded, this constraint is fully lifted and $C_{\mu_{\rm m}}(T \rightarrow \infty) =  Nk_{\rm B}$. The two regimes are separated by a peak associated with the onset of charge excitations.  If the charge density at each lattice site is bounded, as in the case of the MR model of elementary charges, $C_{ \mu_{\rm m}}(T \rightarrow \infty) =  Nk_{\rm B}/2$ and the peak separates the equal low- and high-temperature asymptotes. 
%For the case of the two-dimensional Coulomb gas of purely elementary charges, the low-temperature limit of its specific heat $C_{\rm V}^{\rm elem}$ is the same by the same reasoning. At high temperature, however, the constraint on the electric-field divergence is not lifted but replaced with the constraint $\boldsymbol{\nabla} \cdot \mathbf{E} ({\bf x})=0,\pm q/\epsilon_0 a^2$, so that $C_{\rm V}^{\rm elem}(T \rightarrow \infty) = N k_{\rm B}/2$. 
%Finally, while the HXY specific heat $C_{\rm V}^{\rm HXY}$ again has the same low-temperature behaviour, the statistical dependence between the Coulomb and auxiliary-field components of the HXY lattice electric field further constrains the high-temperature limit: 
%%Finally, while the specific heat of the HXY model $C_{\rm V}^{\rm HXY}$ again has the same low-temperature behaviour, its lattice electric fields are further constrained at high-temperature: 
%the constraint on the electric-field divergence is the same as that imposed on the Coulomb gas of elementary charges, but the magnitude of each of the components of the lattice electric field is also bounded by $J \pi / a$, which places an additional constraint per charge site on the lattice electric fields, so that $C_{\rm V}^{\rm HXY} (T \rightarrow \infty) = 0$. The constraint on the harmonic mode [eq. (\ref{eq:origin-independentEbar})] produces a negligible effect in the thermodynamic limit.
For the HXY model, the charge density is also bounded at all temperatures at each charge site, resulting in one constraint on the lattice electric fields per charge site. In addition to this,  the phase space of the auxiliary field is also bounded, resulting in an additional constraint per charge site on the lattice electric fields in the high-temperature limit. As a consequence of this, the high-temperature limit of the HXY specific heat is zero: $C_{\rm H_{\rm ext}} (T \rightarrow \infty) = 0$. 
The constraint on the harmonic mode [eq. (\ref{eq:origin-independentEbar})] produces a negligible effect in the thermodynamic limit of each model.

This provides an accurate qualitative description of the specific heat of each model through the BKT transition, and explains the strikingly different high-temperature asymptotes of the numerical specific heats %of each model suggested by the data 
presented in fig. \ref{fig:SpecificHeat}.  Recalling from Section \ref{background} that the multi-valued MR model with $q=2\pi$ and $\epsilon_0=1/J$ %and core-energy constants $\epsilon_{\rm c}(m)=0 \, \forall m\in \mathbb{Z}$\footnote{This is an MR model of multi-valued charges.} 
is equivalent to the Villain model with a temperature-independent coupling constant $J$, this difference in the high-temperature specific-heat behaviour displays a clear consequence of the Villain model %with a temperature-independent coupling constant 
not being a pure spin model, i.e., of its local topological defects not being spin vortices. However, given that the specific heat is derived from the second derivative of the logarithm of the partition function with respect to temperature, it follows that the Villain model with a carefully chosen temperature-{\it dependent} coupling constant $J(T)$ could (qualitatively) reproduce the high-temperature specific-heat behaviour of the HXY model. This question is a topic for future work.

Finally, we note that the HXY specific-heat data appears to qualitatively resemble the experimental specific-heat data of thin-film liquid helium-4 presented in fig. 43 of ref.~\cite{Gasparini2008Finite-sizeScalingHelium}.

\subsection{Topological-sector fluctuations}\label{sec:TSF}

Now that we have seen the effect of the statistical coupling of the HXY Coulomb and auxiliary-field components on the specific heat of the system in its high-temperature limit, we turn to a comparison of topological-sector fluctuations in the HXY model with those in the `standard' two-dimensional Coulomb gas statistically decoupled from its auxiliary field. 
As described in Section \ref{background}, the topological sector of the standard Coulomb gas changes when a charge traces a closed path around the torus. It is therefore a measure of charge deconfinement, and its fluctuations were shown to transition from zero to a finite value as the standard Coulomb gas (restricted to local charge dynamics only; global dynamics are described below) %We do not aim to capture the exact physical dynamics; we impose a protocol to evaluate statistical averages of static thermodynamic quantities. Since the quantities are static, we are free to use the term `dynamics' in our discussions.}) 
is heated through the BKT transition~\cite{FBH}.
%Topological-sector fluctuations are \blue{strictly zero in the low-temperature non-ergodic phase of the standard Coulomb gas restricted to local charge dynamics, but dominate its high-temperature ergodic phase}. 
%\blue{In their respective high-temperature phases, the specific heats of the HXY and standard Coulomb systems differ significantly.} In order to understand ergodicity breaking in the HXY model, %-- a more realistic lattice phase model of two-dimensional condensates -- 
%it is therefore pertinent to analyse the effect of the \blue{statistical coupling of the HXY Coulomb and auxiliary-field components on the HXY} topological-sector fluctuations. 
To make this comparison, we note that %, in direct analogy with the standard Coulomb gas, 
the HXY harmonic-mode susceptibility [eq. (\ref{eq:HXYChiEbar})] has a component that only measures the fluctuations in the winding field $\mathbf{\bar{E}}_{{\rm w}}$. %the %which quantifies fluctuations in the topological sector of the system.  } 
%We 
%define the 
We therefore define the HXY winding-field susceptibility %via
\begin{eqnarray}\label{eq:HXYTSF}
\chi_{\rm w}^{\rm H}(L,T):= \beta \epsilon_0^{\rm H}(L,T)L^2\left( \langle \mathbf{\bar{E}}_{\rm w}^2\rangle - \langle \mathbf{\bar{E}}_{\rm w}\rangle^2  \right) .
\end{eqnarray}
%which measures the global topological properties of the system, and is entirely analogous to the function used to quantify topological-sector fluctuations in the standard two-dimensional Coulomb gas~\cite{FBH}. 
This function measures topological-sector fluctuations in the HXY model, in direct analogy with the function used to measure topological-sector fluctuations in the standard Coulomb system~\cite{FBH}. 
%As described in Section \ref{background}, the topological sector of the standard system changes when a charge traces a closed path around the torus. It is therefore a measure of charge deconfinement, and its fluctuations were shown to transition from zero to a finite value as the standard two-dimensional Coulomb gas (restricted to local charge dynamics only\footnote{Local versus global charge dynamics are discussed in the final paragraph of Section \ref{twist}.}) %We do not aim to capture the exact physical dynamics; we impose a protocol to evaluate statistical averages of static thermodynamic quantities. Since the quantities are static, we are free to use the term `dynamics' in our discussions.}) 
%is heated through the BKT transition~\cite{FBH}.

\begin{figure}[t]
\includegraphics[width=\linewidth]{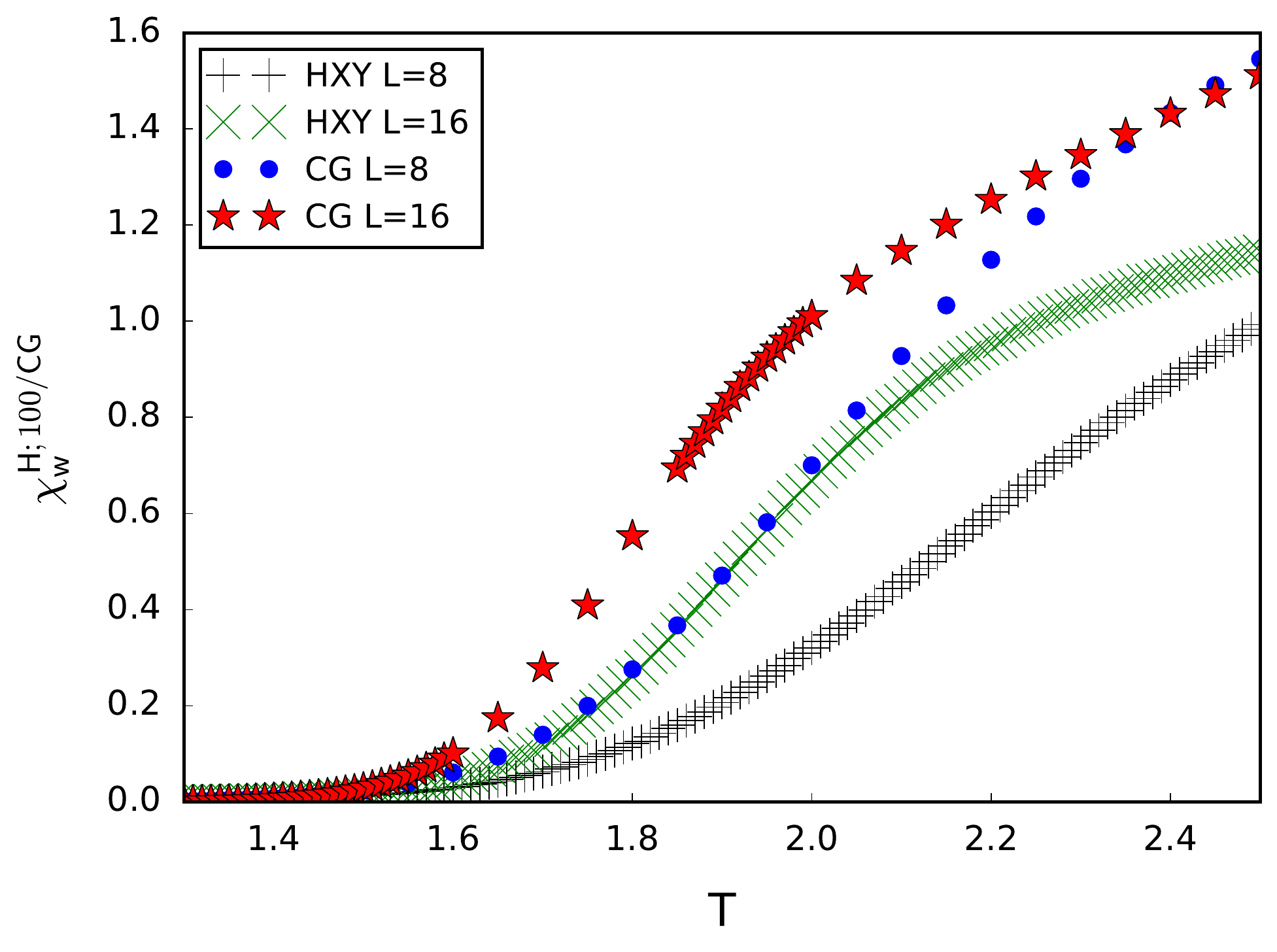}
\caption{Numerical estimates of the HXY winding-field susceptibility $\chi_{{\rm w}}^{{\rm H; }100}$ [whose Fourier series [eq. (\ref{eq:HXYRenormVacuumPerm})] has been cut off at $n=100$] %(with the Fourier series of the temperature-dependent vacuum permittivity of the effective Coulomb system cut off at $n=100$) 
 and the winding-field susceptibility of the two-dimensional Coulomb gas of elementary charges $\chi_{\rm w}^{\rm CG}$ as functions of temperature $T$ for systems of linear size $L=8$ and $16$. %The data tends towards the strict suppression of topological-sector fluctuations in both the effective HXY [with temperature-dependent vacuum permittivity $\epsilon_0^{\rm H}(L,T)$] and standard Coulomb gases  in the low-temperature phase~\cite{FBH}. on. }
For each $L$, $\chi_{{\rm w}}^{{\rm H; } 100}(L,T)\le \chi_{\rm w}^{\rm CG}(L,T) \, \forall T$, with the difference between $\chi_{{\rm w}}^{{\rm H; } 100}(L,T)$ and $\chi_{\rm w}^{\rm CG}(L,T)$ increasing as a function of $T$. %This is a direct consequence of $\epsilon_0^{\rm H}(L,T)$ increasing as a function of $T$ since each $\chi_{\rm w}$ is a measure of charge deconfinement. 
The standard Coulomb gas data and definition of $\chi_{\rm w}^{\rm CG}$ are taken from ref.~\cite{FBH}. We use units such that $J=\epsilon_0=k_{\rm B}=a=1$. Simulation details are described in \ref{app:SimDetails}.} \label{fig:HXYChiw}
\end{figure}
%Fig. \ref{fig:HXYChiw} shows the HXY winding-field susceptibility [with the Fourier series of the temperature-dependent HXY vacuum permittivity $1/ \epsilon_0^{\rm H} (L,T)$ [eq. (\ref{eq:HXYRenormVacuumPerm})] cut off at $n=100$]
Fig. \ref{fig:HXYChiw} shows numerical estimates of the HXY winding-field susceptibility $\chi_{\rm w}^{{\rm H; } 100}$ [whose Fourier series [eq. (\ref{eq:HXYRenormVacuumPerm})] has been cut off at $n=100$] and the winding-field susceptibility of the standard two-dimensional Coulomb gas of elementary charges $\chi_{\rm w}^{\rm CG}$ (as defined in ref.~\cite{FBH}) as functions of temperature $T$ for systems of linear size $L=8$ and $16$. %The data tends towards the strict suppression of topological-sector fluctuations in both the effective HXY [with temperature-dependent vacuum permittivity $\epsilon_0^{\rm H}(L,T)$] and standard Coulomb gases  in the low-temperature phase~\cite{FBH}. 
For each linear system size $L$, $\chi_{{\rm w}}^{{\rm H; } 100}(L,T)\le \chi_{\rm w}^{\rm CG}(L,T) \, \forall T$, with the differences between the magnitudes of the topological-sector fluctuations in the standard Coulomb system and in the HXY model increasing as a function of temperature $T$. As in the case of the two-dimensional Coulomb gas~\cite{FBH}, the data suggests the suppression of HXY topological-sector fluctuations in the low-temperature phase.

Since each $\chi_{\rm w}^{\rm HXY/CG}$ is a measure of charge deconfinement, we interpret the above inequality to be a consequence %we propose that the above inequality could be a direct consequence 
of the HXY charges coupling to their background spin-wave (auxiliary-field) medium, %and that this coupling 
the magnitude of which increases with temperature. This results in relative (with respect to the standard Coulomb gas) decreased HXY charge mobility in the high-temperature phase, but %we still observe 
there remain HXY topological-sector fluctuations with the same qualitative behaviour as those measured in the standard Coulomb gas.

% HXY vacuum permittivity $\epsilon_0^{\rm H}(T)$ of a two-dimensional grand-canonical MR Coulomb gas that describes the HXY system %in the high-temperature phase 
%increasing as a function of temperature $T$: the HXY spin vortices (the charges) couple to the background spin-wave medium (the auxiliary field), which results in relative (with respect to the standard Coulomb gas) HXY charge confinement in the high-temperature phase. %This proposal, however, currently has no analytical basis, hence further work would be required to substantiate this claim. 
%%\blue{Further numerical and analytical work is} required, however, to substantiate this claim. 

Local Coulomb charge dynamics may be complemented with a global dynamics akin to a charge instantaneously tracing a closed path around the torus. This global dynamics instantaneously changes the topological sector of the system. Restricting to configurations of zero charge in the standard two-dimensional Coulomb gas, the probability of a topological-sector fluctuation generated by global charge dynamics is system-size independent~\cite{FBH}. Such topological-sector fluctuations therefore persist down to the zero-temperature limit. 
%This is compared with topological-sector fluctuations in the standard Coulomb system (of non-zero charge) restricted to local charge dynamics being strictly suppressed in its low-temperature phase. 
This is compared with topological-sector fluctuations in the standard Coulomb system (of non-zero charge) restricted to local charge dynamics being strictly suppressed in the low-temperature phase by the logarithmically divergent energy barrier to charge deconfinement. 
It follows that ergodicity is broken between the topological sectors of the standard Coulomb gas restricted to local charge dynamics in the low-temperature phase, where topological-sector fluctuations are ergodically frozen~\cite{FBH}.  %As outlined in \ref{app:globaldynamics}, 
%\blue{The probability of} HXY topological-sector fluctuations generated by global charge dynamics \blue{are} also \blue{system-size independent and therefore also} persist down to the zero-temperature limit. 
In HXY systems restricted to zero charge, topological-sector fluctuations generated by global charge dynamics also persist down to the zero-temperature limit by the same reasoning. 
The data in fig. \ref{fig:HXYChiw} suggests the suppression of HXY topological-sector fluctuations due to local HXY charge dynamics in the low-temperature phase, hence it follows that ergodicity is also broken in the low-temperature phase of the HXY model. 

%\blue{A similar analysis for the 2D-XY model does not produce a system-size independent analogous probability. However, the logarithm of this probability does not diverge in the thermodynamic limit, hence thermodynamic 2D-XY topological-sector fluctuations generated by global charge dynamics persist down to the zero-temperature limit.}

\section{Global spin twists}\label{twist}

\begin{figure*}[t]
\includegraphics[width=\linewidth]{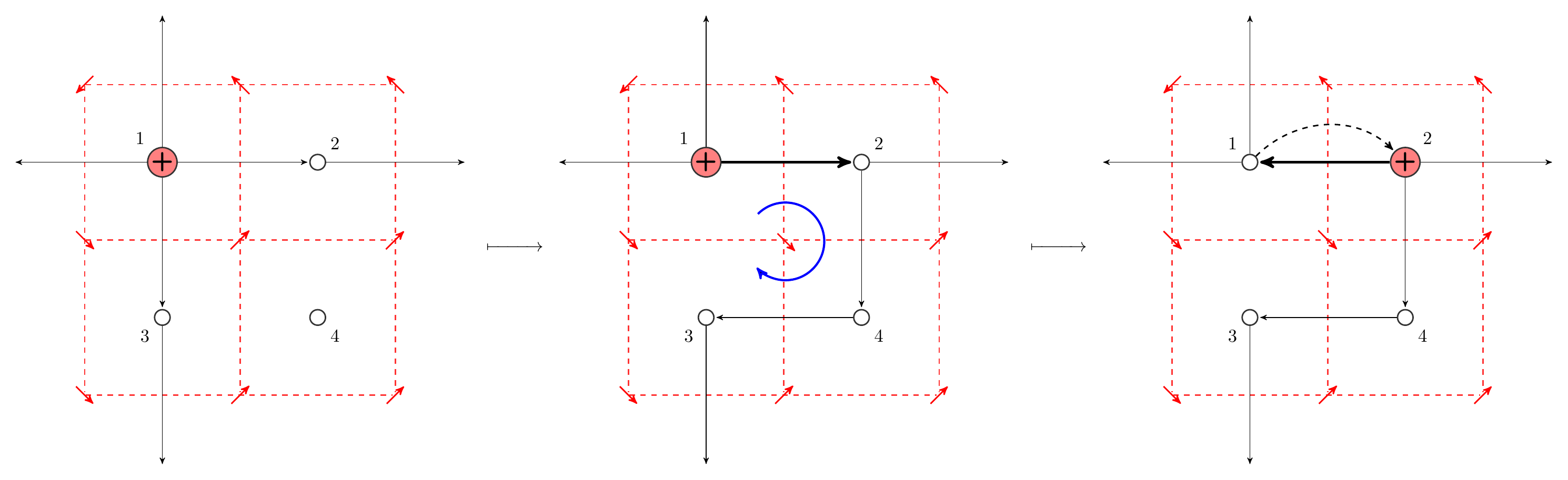}
\caption{A charge hop in the HXY model: The spin at the centre of the diagram has its value decreased by an amount $\pi /2+\omega$ (where we consider the small, positive $\omega$ to ensure that the relevant spin difference leaves the set $\left(-\pi, \, \pi\right]$). This spin rotation is equivalent to an MR auxiliary-field rotation and is followed by an intrinsic modular-symmetry update: the lattice electric field experiences the equivalent of an MR charge-hop update, with $E_{12} +\pi J/2+\omega J \, \mapsto \, E_{12} +\pi J/2 +\omega J - 2\pi J$, where $E_{12}$ denotes the electric flux flowing directly from site $1$ to site $2$ without passing through any other charge sites. The red arrows represent the direction of the spin fixed at each spin site; the intersections of the red dashed lines represent the spin lattice; the large filled red circles represent the (positive) charge/spin-vortex centres; the empty circles represent the unoccupied charge/spin-vortex sites; the black arrows between adjacent charge sites represent the electric flux flowing from one site to the other in the direction of the arrow (with the arrow thickness representing the relative absolute value of the electric flux); the blue arrow represents the direction of the spin rotation.} \label{HXYvortex}
\end{figure*}
In Section \ref{sec:Coulomb}, we established that, while the HXY lattice electric field does not statistically decouple into its Coulomb and auxiliary-field components, %the system does still experience a BKT transition, and 
the system is a spin-model analogue of the MR %electrostatic 
model, and its topological-sector fluctuations have the the same qualitative behaviour as those of the precise 
MR %electrostatic 
model. Here we demonstrate that the coupling between the Coulomb and auxiliary-field components of the HXY lattice electric field transform the ergodic freezing of topological-sector fluctuations into the stronger property of the ergodic exclusion of global HXY spin-twist excitations. 

The microscopic mechanics of (respectively) charge and auxiliary-field fluctuations in the MR  model are depicted in figs. 5 and 6 of ref.~\cite{FBH}. Upon setting $\epsilon_0=1/J$ and $q=2\pi$, the $\varphi$-variable decrease by an amount $\Delta$ shown in fig. 6 of ref.~\cite{FBH} (an MR auxiliary-field rotation) is then equivalent to an HXY spin rotation that does not alter the charge (or spin-vortex) configuration of the system. This corresponds to a rotation of the lattice electric-field flux around the plaquette on which the spin exists by an amount $\Delta J$. An HXY charge hop, however, is a two-step process, and is the result of a particular type of spin rotation. As shown in fig. \ref{HXYvortex}, if a spin rotation results in an absolute spin difference $\delta \varphi_{\nu}^{\mathbf{x}}$ leaving the set $\left(-\pi,\, \pi\right]$, the consequential application of the modular symmetry of the Hamiltonian causes the charge to hop across to the relevant adjacent charge site. This is equivalent to the superposition of a charge hop (fig. 5 of ref.~\cite{FBH}) and an auxiliary-field rotation (fig. 6 of ref.~\cite{FBH}) in the MR  model. The propagation of long-range interactions throughout the HXY model (whose interactions are purely local) then follows by the MR mechanism. We emphasize that the mechanics of the HXY charge hop discussed here are due to the HXY charges being  spin vortices: the composite charge-hop mechanism is special to a system of spin vortices and does not apply to the MR model. %(or Villain model). 

We now reverse our reasoning and analyse the effect on the spins of an HXY charge hopping through the system. As depicted in \ref{app:twist}, an HXY charge that traces a closed path around the torus through an (initially) global-twist-free spin configuration results in a global spin-twist excitation, where the direction through which the spins have twisted is perpendicular to that which the charge has followed in tracing the closed path around the torus: $\bar{E}_{x/y} \mapsto \bar{E}_{x/y} + 2\pi J /L  \Rightarrow   \left[ \boldsymbol{\Delta \bar{\varphi}} \right]_{y/x} \mapsto \left[ \boldsymbol{\Delta \bar{\varphi}} \right]_{y/x} \pm 2\pi a/L$, where $\boldsymbol{\Delta \bar{\varphi}}$ is the harmonic mode of the modular spin-difference field. This is a global HXY spin-twist excitation generated by local charge dynamics. %We note that this excitation occurs in the Villain model, but that it does not correspond to a global spin twist in this case.
To quantify these global twists, we define the low-energy (for a given charge configuration) harmonic mode of the modular spin-difference field to be $\boldsymbol{\Delta \bar{\varphi}}_{\rm p}$ such that
\begin{eqnarray}\label{eq:origin-independentphibar}
\boldsymbol{\Delta \bar{\varphi}}=\boldsymbol{\Delta \bar{\varphi}}_{\rm p}+\frac{2\pi a}{L}\mathbf{t},
\end{eqnarray}
where $\mathbf{t}$ is the integer-valued global spin-twist field, and corresponds to the number of global spin-twist excitations in the system. 
Given that the global spin twists appear due to charges tracing closed paths around the torus, % the global topology associated wit placing the lattice electric fields on the torus, 
it follows from combining the HXY analogue of eq. (\ref{eq:origin-independentEbar}) with eqs. (\ref{emerg_field}) and (\ref{eq:origin-independentphibar}) that $(w_x,w_y)= (t_y,-t_x)$: global HXY spin-twist excitations correspond to non-zero topological sectors of the lattice electric field.  The suppression of topological-sector fluctuations generated by local charge dynamics in the low-temperature phase therefore corresponds to the suppression of global HXY spin-twist excitations generated by local charge dynamics. %(the origin-independent model~\cite{VB,FBH} of the topological sector of the lattice electric field also holds for the 2D-XY model).  %, and hence topological-sector fluctuations in the lattice electric fields of the 2D-XY and HXY models correspond to global spin-twist fluctuations. % and a destruction of the magnetization in finite-size cylindrical and toroidal magnetic films.
We note that $\left| t_{x/y} \right|< L/2a$. %by the same reasonng that non-elementary HXY charges are not possible~\cite{VB}.}

%We now make the connection with the ergodic exclusion of global condensate-phase twists in two-dimensional condensates. 
%In the low-temperature phase of the BKT transition, the global topological sector of the electric field of the standard two-dimensional lattice Coulomb gas restricted to local charge dynamics only is %ergodically frozen 
%constrained to be zero-valued, %its lowest absolute value, 
%which is compared with topological-sector fluctuations generated by global charge dynamics (akin to a charge instantaneously tracing a closed path around the torus) persisting down to low temperatures. 
%It follows that ergodicity is broken between the global topological sectors of the standard Coulomb gas in the low-temperature phase~\cite{FBH}.  As outlined in \ref{app:globaldynamics}, global HXY spin-twist excitations generated by global charge dynamics also persist down to low temperatures. 
In Section \ref{sec:TSF}, we argued that HXY topological-sector fluctuations generated by local charge dynamics are ergodically frozen in the low-temperature phase. 
It follows that global HXY spin-twist excitations generated by local charge dynamics are ergodically %frozen out
excluded in the low-temperature phase. This corresponds to a stronger restriction (on the spins) than the more general ergodic exclusion of non-zero values of the global topological sector of the standard Coulomb gas (i.e., the Villain model with a temperature-independent coupling constant, whose spins are not restricted).

\section{Conclusions}
\label{conclusions}

We presented a lattice electric-field representation of the two-dimensional harmonic XY (HXY) spin model in the context of the Maggs-Rossetto (MR) generalized electrostatic model. 
%With this, 
%We demonstrated that the harmonic mode of the HXY lattice electric field is approximately Coulombically correlated in the low-temperature `Coulomb' phase, but with a temperature-dependent vacuum (electric) permittivity that encodes the coupling of the two-dimensional Coulomb gas of elementary spin vortices to its background spin-wave medium. 
%With this, we performed an analysis of the the purely local HXY spin model as an effective long-range interacting Coulomb gas of elementary spin/condensate-phase vortices in a medium that has a temperature-dependent vacuum electric permittivity. %In mapping from the purely local spin model to the long, we established We presented the mapping b of  in the context of the MR  model. The mapping between the two-dimensional MR and HXY model is through a temperature-dependent HXY vacuum permittivity. 
%\blue{We provided numerical evidence to show that the Coulomb and auxiliary-field components of the HXY lattice electric field energetically decouple and that the system experiences the universal jump of the BKT phase transition.} 
It was then demonstrated that the HXY model is a spin-model analogue of the MR  model, but with a temperature-dependent vacuum permittivity that encodes the coupling of the spin vortices to the background spin-wave medium. 
This coupling results in qualitatively different asymptotic high-temperature specific-heat behaviour in the HXY and MR models; the HXY specific-heat data appears to resemble experimental specific-heat data of thin-film liquid helium-4~\cite{Gasparini2008Finite-sizeScalingHelium}. The topological-sector fluctuations that signal the high-temperature phase of each system, however, display the same qualitative behaviour over the investigated temperature range. 
%The t
%Topological-sector fluctuations %that signal the high-temperature phase of the BKT transition 
%in the standard two-dimensional Coulomb gas then correspond to global HXY spin-twist excitations. 
We also made the observation that the two-dimensional MR  model is equivalent to the Villain model with a temperature-independent coupling constant. 
This is with respect to the grand-canonical lattice electric-field model of multi-valued charges, not the specific algorithm outlined by Maggs and Rossetto~\cite{MR}.

%The effect of spin waves on the effective Coulombic conductivity of the HXY model is similar to the effect of the Jaccard field (or equivalently, Dirac strings and entropic charge) in real~\cite{Ryzhkin,CMS,Jaubert09,magnetricity} and artificial~\cite{kapaklis2014artificalSI} spin ice. The representation of the HXY model presented here could therefore provide a platform with which to better understand these phenomena in spin-ice materials and models. It also seems likely that a frequency regime exists for which the HXY model behaves as an emergent Coulombic conductor under the influence of an ``AC-type'' global twist, similar to that effect seen in spin ice~\cite{Kaiser_SI_Wien}. Also, it is interesting to note that a connection between more complex XY-type models and spin ice has already been established cite{XYspinicePaper}. 

Topological-sector fluctuations in the standard two-dimensional Coulomb gas correspond to global HXY spin-twist excitations.
%The %equivalence 
This correspondence %between global HXY spin-twist excitations and topological-sector fluctuations in the %effective 
%standard two-dimensional %neutral lattice 
%Coulomb gas 
is due to the non-zero topological sectors of the electric field of the %effective 
Coulomb system mapping to global twist configurations in the spin field of the HXY system. As a consequence of this, global HXY spin-twist excitations generated by local HXY charge (or spin-vortex) dynamics are ergodically %frozen out
excluded in the low-temperature phase. %of the BKT transition. %offering a possible pure lattice spin model  for 
This may explain the recently measured~\cite{Shi2016EvidenceCorrelatedDynamicsCuprates} non-ergodic dynamics in layered  cuprates %could be due to the ergodic freezing out of global condensate-phase twists 
at the superconducting transition, where the spins become the phases of the condensate wavefunction. %We note that the wire model of VB~\cite{VB} represents another candidate, but that this system is not a pure lattice spin model in the sense that its spins are defined continuously along the underlying wire-like network, rather than purely at the spin-lattice sites. 
%Future work  -- such as a dynamical analysis of two-dimensional Coulomb physics -- will be required, however, to substantiate this claim.  %The signalling of the high-temperature phase of the BKT transition by topological-sector fluctuations could therefore, in principal, be observable in ultrathin ferromagnetic metallic films~\cite{Ahlberg} or magnetic Langmuir-Blodgett films~\cite{Gayen,Mukhopadhyay}, through measurement of spin correlations 
%\red{I took the Langmuir-Blodgett films part out - could put back in}. 

Further to this, it follows from the Feynman path-integral mapping of ultracold Bosons in a ring lattice that non-zero global spin-twist and topological-sector configurations in the anisotropic 2D-XY model correspond to thermal phase slips in the Bosonic system, and that fluctuations in these 2D-XY configurations potentially signal the dramatic jump in the susceptibility of quantum phase slips at the superfluid -- Mott insulator transition~\cite{Roscilde2016FromQuantum}.

We finally note that the lattice electric-field representation of the HXY model presented here %used to show the equivalence between global spin-twist and topological-sector fluctuations described above 
allows signatures of the magnetic and Coulomb systems to be compared. The finite-size magnetization of the HXY model %and Villain models 
is intimately related to the electric permittivity of the Coulomb gas, and also %, through the mapping of Nelson and Kosterlitz~\cite{Nelson_Kosterlitz}, 
to the superfluid density of condensate films~\cite{Bramwell2015PhaseOrder}. In the future, it will be interesting to explore this connection in more detail.

\ack
It is a pleasure to thank S. T. Banks, G. B. Davies, V. Kaiser, A C Maggs, T Roscilde and A Taroni for valuable discussions. MFF is grateful for financial support from ANR JCJC-2013 ArtiQ. PCWH acknowledges financial support from ANR Grant FISICS. 

\appendix

\section{Microscopic mechanics of global HXY spin-twist excitations generated by local charge dynamics}\label{app:twist}
\begin{figure}[ht]
\includegraphics[width=0.8\linewidth]{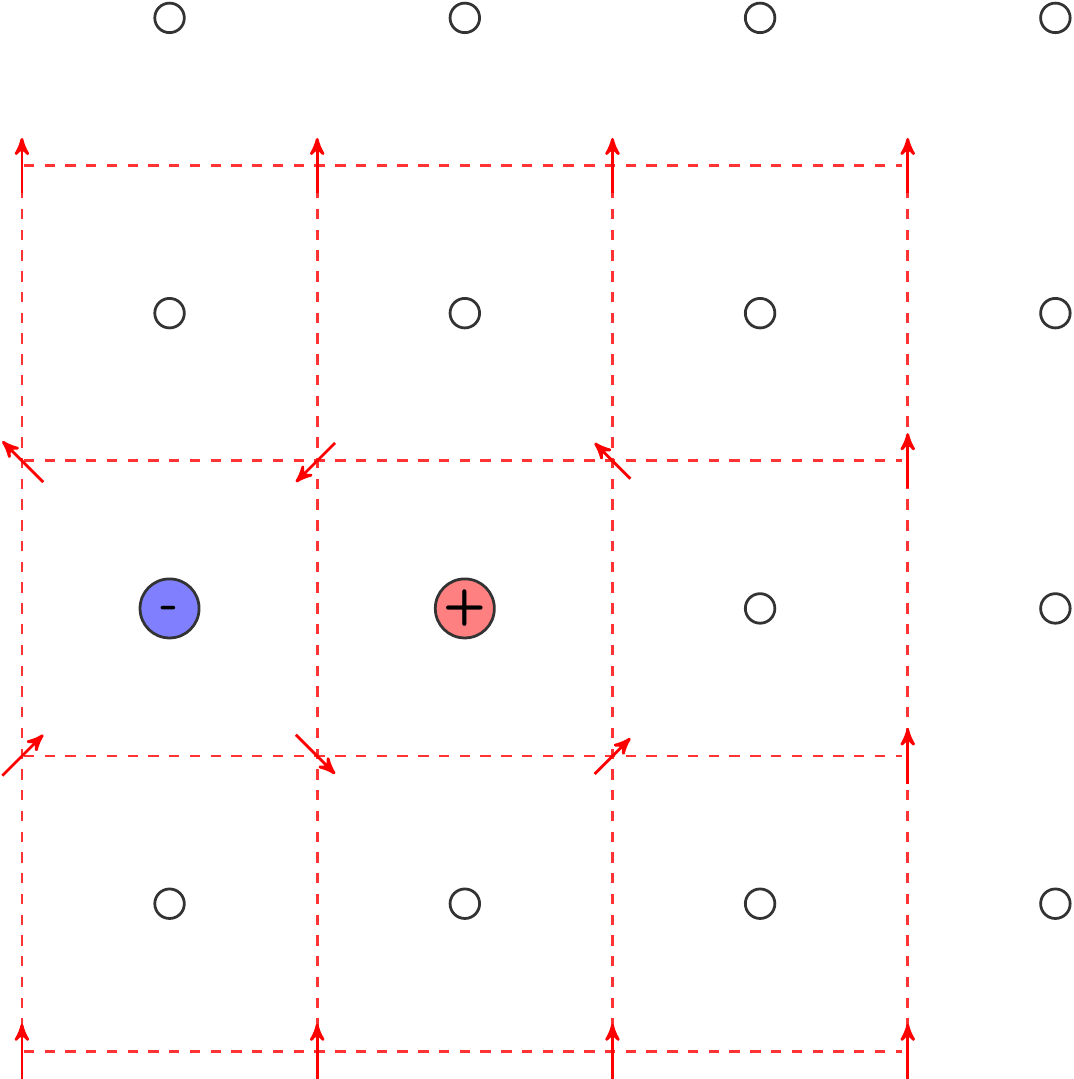}
\caption{A neutral HXY charge pair. The red arrows are the spins; the intersections of the red dashed lines represent the spin lattice; the red/blue circle is a positive/negative charge; the empty circles are empty charge sites.}
\label{fig:twist1}
\end{figure}
\begin{figure}[ht]
\includegraphics[width=0.8\linewidth]{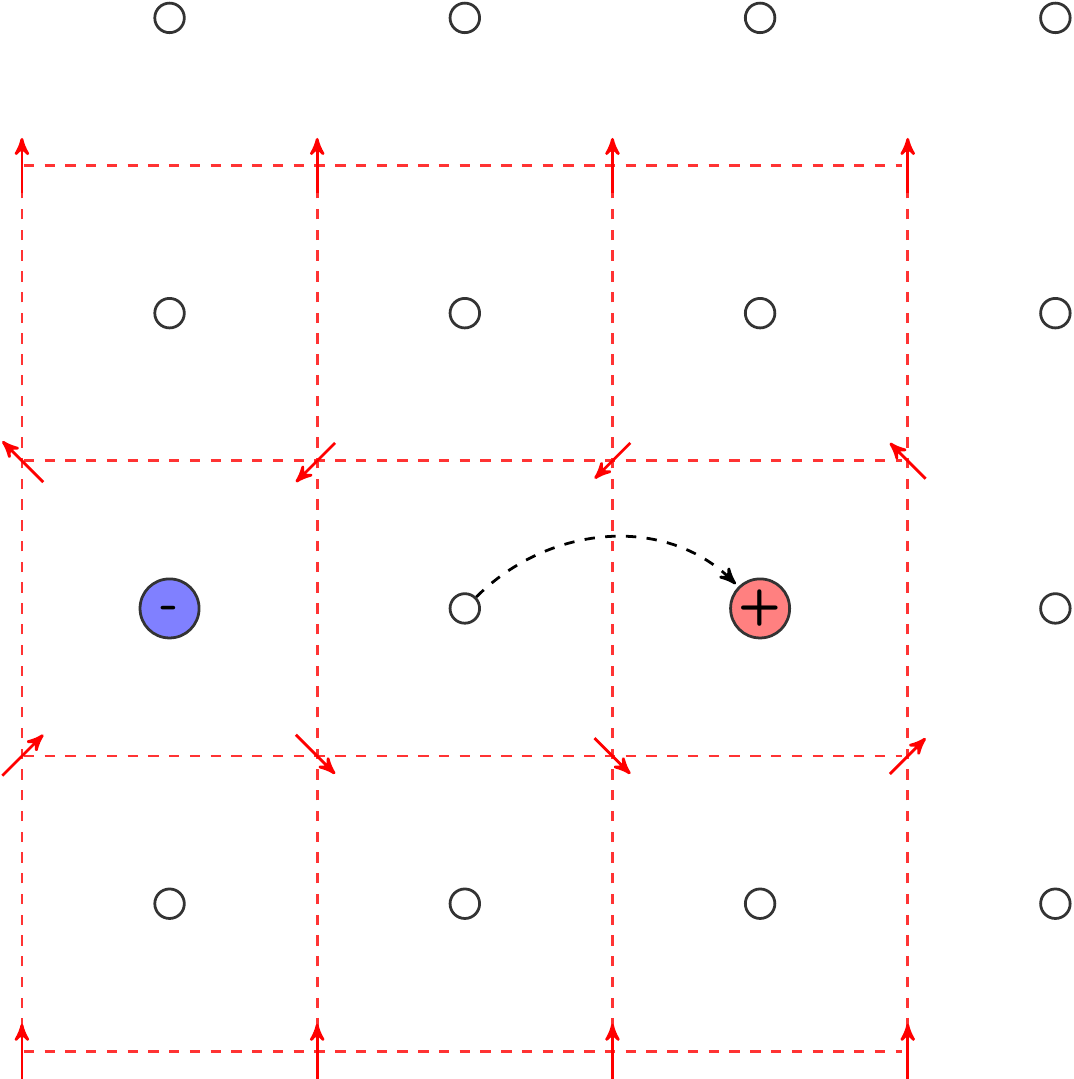}
\caption{The positive charge hops in the positive $x$ direction. The curly dashed black arrow represents a charge hop.}
\label{fig:twist2}
\end{figure}
\begin{figure}[ht]
\includegraphics[width=0.8\linewidth]{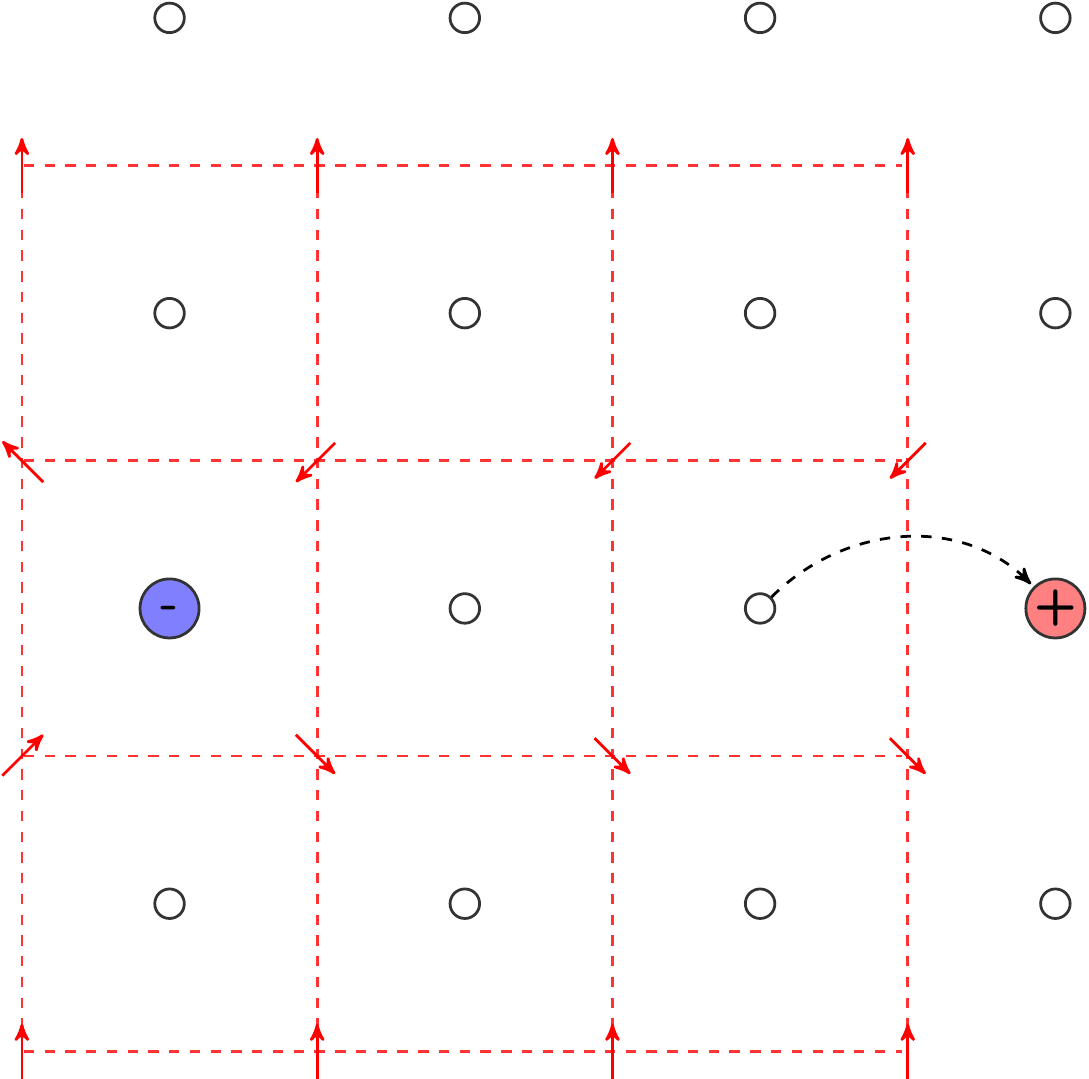}
\caption{The positive charge hops in the positive $x$ direction again.}
\label{fig:twist3}
\end{figure}
\begin{figure}[ht]
\includegraphics[width=0.9\linewidth]{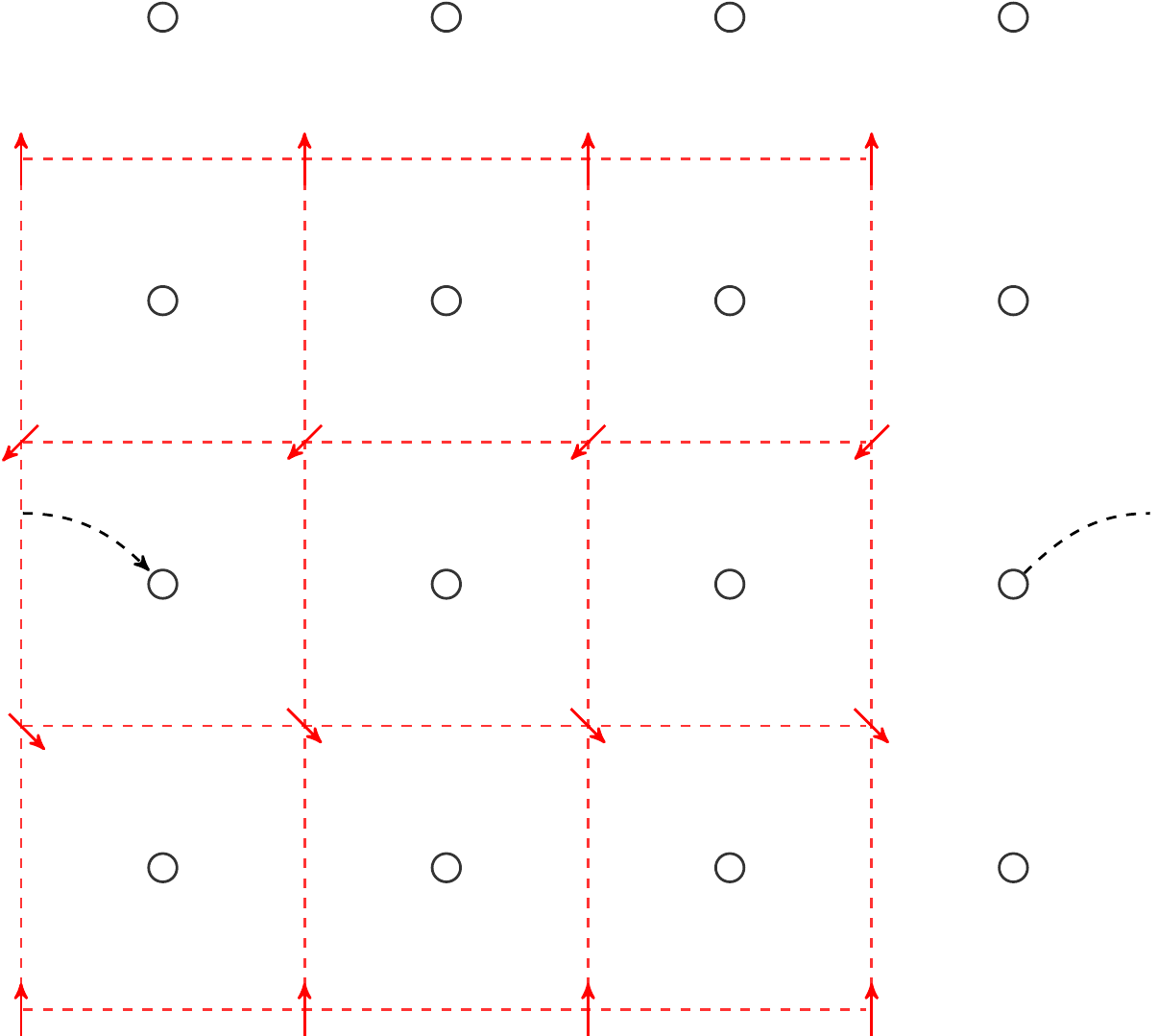}
\caption{The positive charge hops in the positive $x$ direction once more, annihilating the negative charge.}
\label{fig:twist4}
\end{figure}
\begin{figure}[ht]
\includegraphics[width=0.9\linewidth]{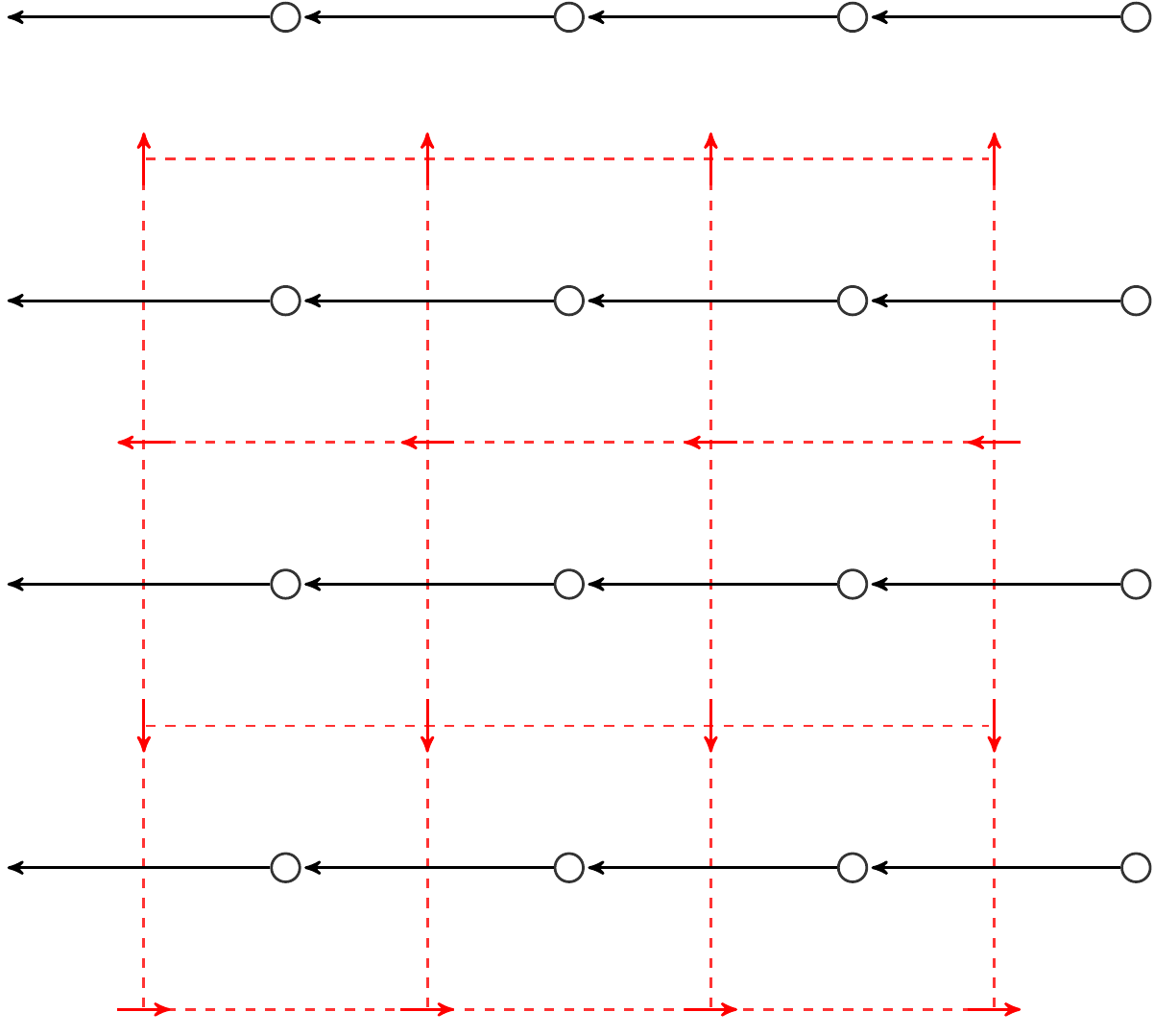}
\caption{The minimum-energy field configuration corresponding to fig. \ref{fig:twist4} (up to a global rotation). The electric-field representation is superimposed in black.}
\label{fig:twist5}
\end{figure}
Figs. \ref{fig:twist1} to \ref{fig:twist5} depict the creation of a global HXY spin twist by local HXY charge dynamics. PBCs are applied in both directions. Fig. \ref{fig:twist1} shows a neutral charge pair that has been created out of the vacuum. In figs. \ref{fig:twist2} to \ref{fig:twist4}, the positive charge traces a closed path around the torus in the
$x$ direction until it annihilates the negative charge in fig. \ref{fig:twist4}. The remnant minimum-energy field configuration is then shown in fig. \ref{fig:twist5} (up to a global rotation). The electric-field representation is superimposed in black. The final spin configuration is a global spin twist $t_y=-1$ in the $y$ direction, while the final electric-field configuration has $w_x=-1$, which corresponds to a non-zero global topological sector.

\section{Simulation details}\label{app:SimDetails}

The data in fig. \ref{fig_energy} were generated from one run of a Metropolis Monte Carlo HXY simulation of $10^5$ sweeps per spin and $10000$ quench sweeps per measurement. The remaining data sets presented in this paper were averaged over multiple Metropolis Monte Carlo runs of $10^6$ spin-update/charge-hop sweeps per lattice site for the HXY/MR models. 5 auxiliary-field sweeps were performed per charge-hop sweep for the MR data. 
The data in fig. \ref{fig:HXYHelicity} were averaged over 64 runs for the systems of linear size $L=8$, $16$ and $64$, and 48 runs for the system of linear size $L=32$. The MR data in fig. \ref{fig:SpecificHeat} were averaged over 64 runs for all data points except $T=1.15-1.6$ and $T=1.85-2.0$ for the elementary MR model, which were averaged over 256 runs; the HXY data in fig. \ref{fig:SpecificHeat} were averaged over 32 runs. The HXY data in fig. \ref{fig:HXYChiw} were averaged over 64 runs; the MR data in fig. \ref{fig:HXYChiw} were averaged 256 runs.

%\subsection{The 2D-XY model}

%Similarly, the constrained 2D-XY winding-field susceptibility $\chi_{\rm w}^{\rm XY,g}$ becomes
%\begin{eqnarray}
%\chi_{\rm w}^{\rm XY,g} &\simeq&  \frac{16 \pi^2 \beta J\exp \left[\beta JN -2\pi^2 \beta J + O\left(\frac{1}{N}\right) \right]}{e^{\beta J N}+ 4\exp \left[ \beta JN -2\pi^2 \beta J + O\left(\frac{1}{N}\right) \right]} \nonumber\\
%&\rightarrow & 16\pi^2 \beta J  \exp \left( -2\pi^2 \beta J\right) 
%\end{eqnarray}
%as $N\rightarrow \infty$. Hence, 2D-XY global spin-twist fluctuations generated by global charge dynamics persist down to low temperatures in the thermodynamic limit.

\section*{References}

\bibliography{Bibliography_new}{}

\end{document}